\documentclass[%
twocolumn,
aip,
floatfix,
 amsmath,amssymb,
reprint,%
]{revtex4-1}

\usepackage{graphicx}
\usepackage{dcolumn}
\usepackage{bm}
\usepackage{siunitx}
\usepackage[utf8]{inputenc}
\usepackage[T1]{fontenc}
\usepackage{mathptmx}
\usepackage{etoolbox}
\usepackage{float}
\usepackage{xcolor}
\usepackage{natbib}
\usepackage{microtype}
\usepackage{soul}
\setstcolor{blue}

\makeatletter
\def\@email#1#2{%
 \endgroup
 \patchcmd{\titleblock@produce}
  {\frontmatter@RRAPformat}
  {\frontmatter@RRAPformat{\produce@RRAP{*#1\href{mailto:#2}{#2}}}\frontmatter@RRAPformat}
  {}{}
}%
\makeatother
\begin{document}

\preprint{AIP/123-QED}

\title{Lattice Boltzmann simulation of deformable fluid-filled bodies: progress and perspectives}

	\author{Danilo P. F. Silva}
	\affiliation{Centro de F\'isica Te\'orica e Computacional, Faculdade de Ciências, Universidade de Lisboa,
		P-1749-016 Lisboa, Portugal}
    \affiliation{Departamento de F\'{\i}sica, Faculdade
		de Ci\^{e}ncias, Universidade de Lisboa, P-1749-016 Lisboa, Portugal}
		
		 \author{Rodrigo C. V. Coelho}
    \email{dpsilva@fc.ul.pt, rcvcoelho@fc.ul.pt}
	\affiliation{Centro de F\'isica Te\'orica e Computacional, Faculdade de Ciências, Universidade de Lisboa,
		P-1749-016 Lisboa, Portugal}
    \affiliation{Departamento de F\'{\i}sica, Faculdade
		de Ci\^{e}ncias, Universidade de Lisboa, P-1749-016 Lisboa, Portugal}

  \author{Ignacio Pagonabarraga}
   \affiliation{Departament de Física de la Matèria Condensada, Universitat de Barcelona,
Carrer de Martí Franqués 1, 08028 Barcelona, Spain}
  \affiliation{Universitat de Barcelona Institute of Complex Systems (UBICS),
Universitat de Barcelona, 08028 Barcelona, Spain}

  \author{Sauro Succi}
   \affiliation{Center for Life Nano Science at La Sapienza, Istituto Italiano di Tecnologia, 295 Viale Regina Elena, I/00161 Roma, Italy}
  \affiliation{Harvard Institute for Applied Computational Science, Cambridge, MA 02138, USA}
	
	\author{Margarida M. Telo da Gama}
	\affiliation{Centro de F\'isica Te\'orica e Computacional, Faculdade de Ciências, Universidade de Lisboa,
		P-1749-016 Lisboa, Portugal}
    \affiliation{Departamento de F\'{\i}sica, Faculdade
		de Ci\^{e}ncias, Universidade de Lisboa, P-1749-016 Lisboa, Portugal}
	
	\author{Nuno A. M. Ara\'ujo}
	\affiliation{Centro de F\'isica Te\'orica e Computacional, Faculdade de Ciências, Universidade de Lisboa,
		P-1749-016 Lisboa, Portugal}
    \affiliation{Departamento de F\'{\i}sica, Faculdade
		de Ci\^{e}ncias, Universidade de Lisboa, P-1749-016 Lisboa, Portugal}

\date{\today}

\begin{abstract}
With the rapid development of studies involving droplet microfluidics, drug delivery, cell detection, and microparticle synthesis, among others, many scientists have invested significant efforts to model the flow of these fluid-filled bodies. Motivated by the intricate coupling between hydrodynamics and the interactions of fluid-filled bodies, several methods have been developed. The objective of this review is to present a compact foundation of the methods used in the literature in the context of lattice Boltzmann methods. For hydrodynamics, we focus on the lattice-Boltzmann method due to its specific ability to treat time- and spatial-dependent boundary conditions and to incorporate new physical models in a computationally efficient way. We split the existing methods into two groups with regard to the interfacial boundary: {\it fluid-structure} and {\it fluid-fluid methods}. The fluid-structure methods are characterised by the coupling between fluid dynamics and mechanics of the flowing body, often used in applications involving membranes and similar flexible solid boundaries. We further divide fluid-structure-based methods into two subcategories, those which treat the fluid-structure boundary as a continuum medium and those that treat it as a discrete collection of individual springs and particles. Next, we discuss the fluid-fluid methods, particularly useful for the simulations of fluid-fluid interfaces. We focus on models for immiscible droplets and their interaction in a suspending fluid and describe benchmark tests to validate the models for fluid-filled bodies.
\end{abstract}

\maketitle


\section{Introduction}

The many-body dynamics of deformable objects, passive and active alike, in
confined fluid flows is a central theme of modern soft matter research, with 
many applications in science, engineering, and medicine. 
From a fundamental perspective, the main challenge relates to the 
self-consistent interplay between external degrees of freedom, positions 
and momenta, and the internal ones describing the shape changes of the deformable body. 
Such an interplay is expected to spawn new dynamic regimes that are
simply inaccessible to rigid particles. Likewise, such new dynamic regimes
are expected to lead to new states of soft matter, hence
potentially new materials.

In this review, we shall focus on the transport of deformable fluid-filled 
bodies, an important problem with various applications, from oil 
to pharmaceutical and food-processing industries
~\cite{Anna_Mayer_2006, Liu_Zhang_2011,Clausen_Aidun_2010,Sui_Chew_Low_2007}. 
Familiar examples include paints \cite{gennes2004capillarity,israelachvili2022surface,qi2020dynamic}, milk \cite{truong2014effect,dickinson2012emulsion,mcclements2023modeling}, and blood \cite{fedosov2014deformation, dupire2012full}, where the internal constituents 
are ink droplets, fat droplets, and cells, respectively, and all deform when 
subjected to mechanical stresses, such as the ones resulting from the flow 
of the surrounding fluid. 
An analytical approach proves challenging due to the strong coupling between 
the fluid and the deformable particles. 
This interaction is reciprocal, for the fluid affects the shape of the particle 
due to hydrodynamic forces and the deformation in turn changes the boundaries of the 
fluid flow. While still facing numerous challenges, computational methods 
have become an indispensable alternative tool to study 
these systems~\cite{aidun_lattice-boltzmann_2010,BENZI1992145,Ye_Pan_Huang_Liu_2019,Zhu_Zhou_Yang_Yu_2008}. 

\begin{figure}
\center
\includegraphics[width=0.48\textwidth]{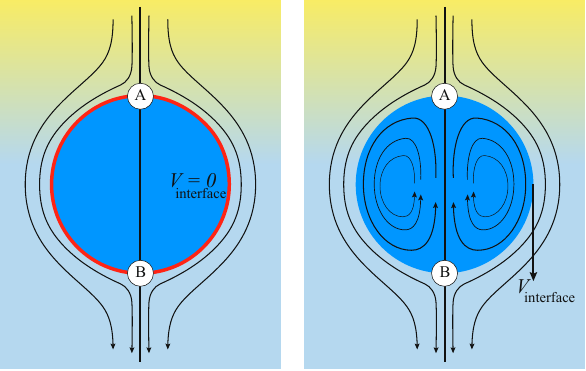}
\caption{ Comparison of flow around a fluid-solid (left) and a fluid-fluid (right) interface. On the left, the red line represents a fluid-solid interface such as a membrane on a rigid capsule. There is no internal circulation because the velocity at the interface is zero (considering no-slip boundary conditions). On the right, we see internal circulation due to the continuity of the velocity across the fluid-fluid interface. Points A and B indicate stagnation points and the colour gradient of the external fluid represents a pressure gradient which drives the flow. Both images are in the capsule reference frame (fluid velocity relative to the capsule/droplet). The internal fluid in both images is blue. Black arrows represent the fluid flow.}
\label{fig:dropletvscapsule}
\end{figure}

Numerical models for the interplay between the macroscopic fluid properties 
and its internal deformable constituents usually combine computational fluid dynamics (CFD) with 
computational mechanics \cite{Gao_Hu_2009,kruger_efficient_2011,Sugiyama_Ii_Takeuchi_Takagi_Matsumoto_2010}. 
This is particularly true for the simulation of fluid-filled bodies with solid boundaries. 
Deformability alters the flow, hence it affects the way energy is dissipated 
within the flow, resulting in strong rheological 
and mechanical nonlinearities, as signalled by the crossover from 
Newtonian to shear-thinning regimes, such as in most emulsions.
A fluid-filled body consists of a body, usually spherical, with a fluid 
core (see Fig. \ref{fig:dropletvscapsule}). 
Common examples include droplets and capsules. 
The latter is a liquid droplet enclosed by a thin membrane. 
These bodies are often surrounded by a fluid environment and consequently, they 
exhibit fluid-fluid boundaries (droplets) or fluid-solid boundaries (capsules).

The modelling approaches for fluid-filled bodies can be divided into two groups, based 
on the boundary between the fluid-filled body and its environment. 
These are fluid-structure and fluid-fluid methods. 
The former consists of modelling a fluid-solid boundary, where the fluid and a deformable solid interact with each other. 
The second consists of modelling the interactions between two fluids, usually separated 
by a fluid-fluid interface. 
The nature of the boundaries requires computational methods to 
describe both the fluid inside and outside of the particles. 
Furthermore, tracking the interface is often required, which can make the overall 
process computationally expensive. 
In this review, we focus on cases where the internal and external fluids are 
both Newtonian.  
We point out that for industrial applications, capsules can have solid 
cores (even multiple small solid cores) enclosed by a membrane; however this 
will not be addressed in this review \cite{Gao_Chen_2019}.

Computational methods can often illuminate regions of parameter space 
not easily accessible by experiments or analytical methods. 
Several methods can be used to model the fluid flow. 
Here, we focus on the lattice Boltzmann method (LBM), as it provides a 
versatile way to introduce fluid-fluid interactions and to couple fluid-solid 
boundaries as well. 
Various methods (continuum, discrete, mesoscopic) have been developed to 
simulate droplets, cells, capsules and so on, and we refer to existing 
literature on this \cite{Hou_Wang_Layton_2012,Schiller_Kruger_Henrich_2018}. 
We restrict ourselves to those methods that employ or pertain to LBM. 
Previous reviews and books~\cite{Karimnejad2022, OUP18, KRUGER} have covered some fluid-solid 
and fluid-fluid methods in a different context. 
For instance, Karimnejad et. al.~\cite{Karimnejad2022} have reviewed fluid-structure (FS) methods for rigid particles. 
Here our focus is on methods for deformable particles in fluids. 
We present a comprehensive review of those methods and compare them.

Finally, we wish to underline that the methods described in this review straddle 
across the passive versus active soft matter divide, just because, once the coupling
between internal and external degrees of freedom is accounted for, such divide 
largely blurs out~\cite{PhysRevLett.130.130002}. The methods described here may be used to investigate problems in the active matter systems such as the emergent nematic order in epithelial cells, which share many similarities with the emulsions of Sec.~\ref{sec_fluid_fluid}, and the shape changes in bacteria, which become more elongated when swarming~\cite{D2SM00988A,Beer2019}.

The review is organised as follows. 
The LBM is summarised in Sec.~\ref{sec:LBM}. In Sec.~\ref{sec:FSI}, we review 
fluid-structure methods which often use the immersed boundary method. 
We highlight the differences and similarities between the continuum and the 
discrete approaches. 
In Sec.~\ref{sec_fluid_fluid}, we review the various multicomponent 
methods for immiscible droplets. In Sec.~\ref{sec:benchmarks}, we provide benchmarks 
in 2D and 3D for hydrostatic and hydrodynamic conditions. 
In Sec.~\ref{sec:conclusions}, we conclude by summarising our observations 
and make remarks on open questions and future challenges.

\section{Lattice Boltzmann method}

\label{sec:LBM}
In recent years the LBM has become a useful tool in solving the hydrodynamics of soft matter systems \cite{Dunweg_lattice_2009}. The fluid flow is described by the Navier-Stokes (NS) equations, 
\begin{equation}
\begin{aligned}
&\rho\left(\frac{\partial \boldsymbol{u}}{\partial t}+(\boldsymbol{u} \cdot \nabla) \boldsymbol{u}\right)=-\nabla p+\mu \nabla^{2} \boldsymbol{u}+ \boldsymbol{g}, \\
&\nabla \cdot \boldsymbol{u}=0,
\end{aligned}
\label{eq:NS_equations}
\end{equation}
where $\boldsymbol{u}$ is the fluid velocity, $t$ is time, $p$ is the pressure field, $\mu$ is the dynamic viscosity, $\rho$ is the density and $\boldsymbol{g}$ is the body force. LBM however does not solve the NSE directly. Instead, it is a mesoscopic method which recovers the results of the NS equations. As opposed to molecular dynamics, which solves the dynamics of individual molecules/atoms, in LBM a fluid ``particle'' consists of a group of molecules and consequently is much larger than the individual scale of molecules. Information about the fluid is contained in the lattice nodes and can only move in specific directions reducing the number of degrees of freedom. Although it originates from Lattice Gas Automata \cite{frisch_lattice-gas_1986}, it is now considered a separate method. Several works have shown the potential of LBM to solve fluid-related problems~\cite{qian_lattice_1992,he_lattice_1997,gunstensen_lattice_1991,chen_lattice_1998,aidun_lattice-boltzmann_2010,succi_lattice_2001}. It is an efficient algorithm particularly popular for Stokes flow, systems with complex solid boundaries \cite{ladd_numerical_1994},  multiphase flow \cite{he1999a}, hydrodynamic dispersion~\cite{PhysRevA.45.5771} and convection \cite{he_novel_1998}. 

\subsection{Lattice BGK model}
\label{subsec:bgkmodel}
The  Boltzmann equation that contains the fluid particle information reads
\begin{equation}
\frac{\partial f}{\partial t}+\boldsymbol{v} \cdot \nabla f+\boldsymbol{a} \cdot \nabla_{\boldsymbol{v}} f=\Omega(f)
\label{eq:DBE}
\end{equation}
where $f=f(\boldsymbol{x}, \boldsymbol{v}, t)$ is the distribution function for a fluid particle in phase space, having microscopic velocity, $\boldsymbol{v}$, at time $t$, at position $\boldsymbol{x}$, where $\boldsymbol{a}$ is the acceleration and $\Omega(f)$ is the collision operator which describes the interaction between molecules. Different LBM formulations will have different collision operators. The simplest one is the Bhatnagar-Gross-Krook (BGK) operator given by
\begin{equation}
\Omega(f)=-\frac{f(\boldsymbol{x}, \boldsymbol{v}, t)-f^{e q}(\boldsymbol{x}, \boldsymbol{v}, t)}{\tau},
\end{equation}
where $\tau$ is the relaxation time which drives the relaxation to the equilibrium distribution $f^{e q}(\boldsymbol{x}, \boldsymbol{v}, t)$. The relaxation time is directly related to the viscosity of the fluid since the fluid kinematic viscosity $\nu$ is given by
\begin{equation}
\nu=c_{s}^{2}\left(\tau-\frac{\Delta t}{2}\right),
\label{eqn:viscosity_ls}
\end{equation}
where $c_s$ is the  speed of sound, $\tau$ is the single relaxation time and $\Delta t$ is the time step. 
It should be stated, however, that connecting the kinematic viscosity and 
the speed of sound is shown to be valid for low Mach number ($< 0.3$) 
and nearly incompressible flow.

\begin{figure}
\center
\includegraphics[width=0.29\textwidth]{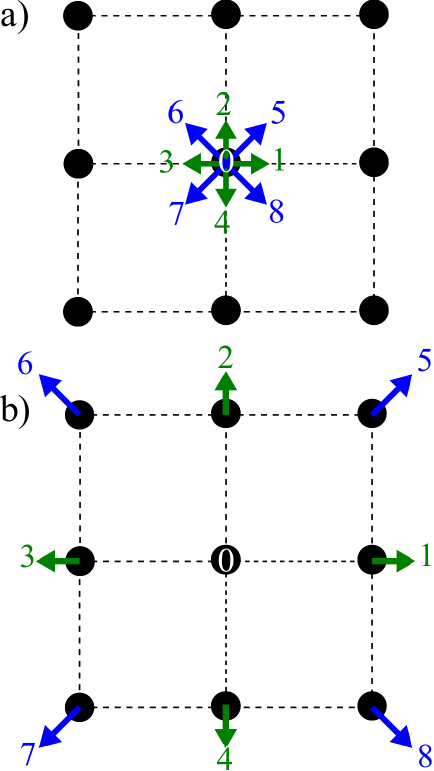}
\caption{Two main steps of the LBM: (a) collision and (b) streaming using a D2Q9 lattice. The collision is described by the distribution functions going from one node after the streaming step propagates them to the neighbouring nodes according to the direction. The arrows represent the nine distribution functions.}
\label{fig:streaming_collision}
\end{figure}

The equilibrium distribution function is related to the macroscopic 
fluid velocity $\boldsymbol{u}$ and density $\rho$
\begin{equation}
f^{eq} (\mathbf{x}, \vert \boldsymbol{v}\vert, t)=\frac{\rho}{(2 \pi \theta)^{D / 2}} \exp \left[-\frac{(\boldsymbol{v}-\boldsymbol{u})^2}{2 \theta}\right],
\end{equation}
where $\theta$ is the normalised temperature which is usually assumed to be the unity and $D$ is the spatial dimension. Note that $\rho$, $\boldsymbol{u}$ and $\theta$ depend on time and space in general. The above equation can be expanded using the Hermite series to the second order as
\begin{equation}
f_{\alpha}^{eq}\left(\boldsymbol{x}_{i}, t\right)=w_{\alpha} \rho\left(1+\frac{\boldsymbol{\xi}_{\alpha} \cdot \boldsymbol{u}}{c_{s}^{2}}+\frac{\left(\boldsymbol{\xi}_{\alpha} \cdot \boldsymbol{u}\right)^{2}}{2 c_{s}^{4}}-\frac{\boldsymbol{u}^{2}}{2 c_{s}^{2}}\right),
\label{eqn:eq_distribution}
\end{equation}
where the subscript $\alpha$ represents the discrete lattice speed (see Fig. \ref{fig:streaming_collision}) , $\boldsymbol{\xi}_{\alpha}$ is 
the microscopic velocity vector, $w_{\alpha}$ 
is the weight of the lattice and, $i$ is the index of lattice sites. $f_{\alpha}^{eq}\left(\boldsymbol{x}_{i}, 
t\right)$ is the distribution at equilibrium at position $\boldsymbol{x}_{i}$ and time $t$. 
Both $\rho$ and $\boldsymbol{u}$ are a function of position $\boldsymbol{x}_{i}$ and time $t$. The parameters $\boldsymbol{\xi}_{\alpha}$ and $w_{\alpha}$ depend on the lattice chosen for the discretization. The velocity is discretized into structured lattice velocity (1D, 2D or 3D) vectors. For example, the D1Q3 lattice represents a one-dimensional chain with three vectors per node. Usually, more lattice vectors mean increased precision but also increased computational effort. Popular lattices include the D2Q9 (see Fig. \ref{fig:streaming_collision}) and D3Q19. Different lattices have different speeds of sound $c_s$ and weights $w_\alpha$ for each vector \cite{kruger_lattice_2017}.

The final discrete lattice Boltzmann equation (with the BGK operator) is,
\begin{multline}
\underbrace{f_{\alpha}\left(\boldsymbol{x}_{i}+\boldsymbol{\xi}_{\alpha} \Delta t, t+\Delta t\right)}_{\text {streaming }}=\\\underbrace{f_{\alpha}\left(\boldsymbol{x}_{i}, t\right)-\frac{\Delta t}{\tau}\left[f_{\alpha}\left(\boldsymbol{x}_{i}, t\right)-f_{\alpha}^{eq}\left(\boldsymbol{x}_{i}, t\right)\right]+\mathcal{F}_{\alpha}\Delta t}_{\text {collision }},
\label{eq:discrete_LBE}
\end{multline}
where $\mathcal{F}_{\alpha}$ is the forcing term which includes both external forces and intra/intermolecular fluid forces. There are different force schemes in LBM used to implement these terms. 
Several articles and reviews have been written about the advantages/disadvantages 
of each one (see Ref. \cite{bawazeer_critical_2021,sun_evaluation_2012,chen_critical_2014}). 
A popular one, the Guo forcing scheme \cite{guo2002a}, is given by 
\begin{equation}
\mathcal{F}_{ \alpha}\Delta t=\left(1-\frac{1}{2 \tau{ }}\right) w_{ \alpha}\left[\frac{\boldsymbol{\xi }_{ \alpha}-\boldsymbol{u}}{c_{s}^{2}}+\frac{\left(\boldsymbol{\xi }_{ \alpha} \cdot \boldsymbol{u}\right) \boldsymbol{\xi}_{ \alpha}}{c_{s}^{4}}\right] \cdot \boldsymbol{F},
\label{eq:guo_force}
\end{equation}
where $\boldsymbol{F}$ is the force. There are two main steps in Eq. \ref{eq:discrete_LBE}: collision and streaming (see Fig.~\ref{fig:streaming_collision}). The information for both steps comes from the nearest neighbours (usually within the first belt) of the respective fluid node. This makes the method very adaptable to parallel computing such as with graphical processing units (GPUs), when compared to other conventional methods.

The macroscopic density and velocity are calculated through
\begin{equation}
\begin{aligned}
\rho(\boldsymbol{x}, t) &=\sum_{\alpha} f_{\alpha}(\boldsymbol{x}, t), \\
\rho \boldsymbol{u}(\boldsymbol{x}, t) &=\sum_{\alpha} \boldsymbol{\xi}_{\alpha} f_{\alpha}(\boldsymbol{x}, t).
\end{aligned}
\label{eq:macrosocpicvariables}
\end{equation}

In this review, we focus on the LBM to model the hydrodynamics. The LBM is particularly well suited to multiphase flow, particularly drops, bubbles, and emulsions. Some of the commonly quoted strengths
and weaknesses of the LBM method are discussed, for example, in Ref.~\cite{kruger_lattice_2017,huang_multiphase_2015}. We also focus on the single relaxation time collision operator due to its simplicity, but we emphasise that more sophisticated ones are used to mitigate spurious effects and may offer enhanced stability, like the MRT collision operator~\cite{PhysRevE.78.016701, PhysRevE.61.6546} and the entropic LBM~\cite{doi:10.1098/rspa.2000.0689, PhysRevLett.114.174502}. 

\section{Fluid-Structure based methods}
\label{sec:FSI}
\begin{figure*}
\center
\includegraphics[width=1.000\textwidth]{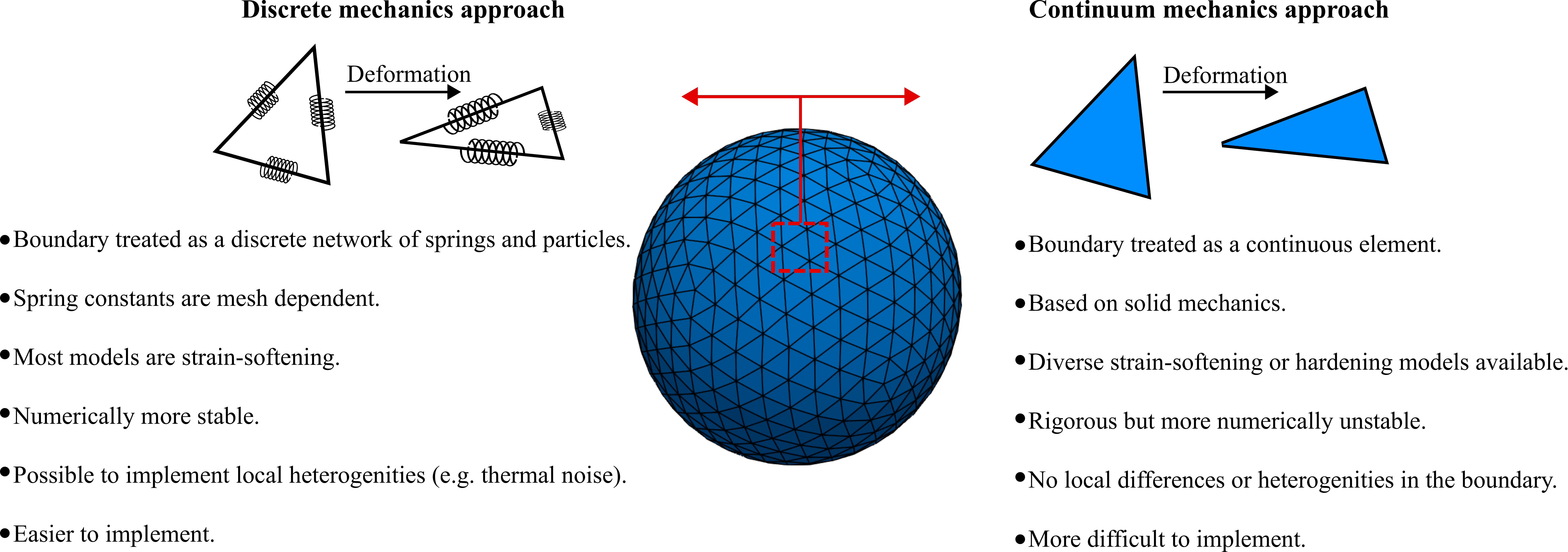}
\caption{Discrete mechanics approach (left) vs continuum mechanics approach (right). On the discrete mechanics approach the dynamics of the membrane is given by a connected network of springs. On the continuum mechanics approach the membrane is treated as a surface i.e. a continuum medium. The advantages and disadvantages of each approach are highlighted.}
\label{fig:continuumvsdiscrete}
\end{figure*}

At the microscale, there is a plethora of particles which are bound by membranes. Examples include living cells, artificial capsules (e.g. polymerised membranes) and vesicles (e.g. lipid bilayer membranes). An example of biological membrane-bound particles is red blood cells (RBCs). We highlight that variations of the models presented here have been used for soft objects like microgels \cite{lei_transport_2019} and capsules \cite{zhang_immersed_2007}. While these bodies can suffer strong deformations, for the majority of them volume is relatively unchanged. Consequently, researchers have been interested in the transport of soft matter and other complex fluids flowing at low Reynolds numbers. In order to model biological and synthetic membrane-bound particles, one should consider the membrane elastic, bending and viscous properties along with the inner and outer fluid viscosity. Depending on what is being modelled a combination of the aforementioned properties is necessary to accurately capture the physics. Furthermore, the meshing of the membrane is required, particularly for the methods described in this section. In 2D, the mesh consists of line segments and 3D of surfaces. In general, there are two ways to approach this problem. These approaches depend on whether the boundary (in most cases a membrane) is treated as a continuum medium or as a discrete one. We show in Fig. \ref{fig:continuumvsdiscrete} an illustration of the two different approaches. 

The first approach is a continuum one where the membrane is treated as a 2D continuum surface in 3D space. One simulates the solid mechanics of the membrane using appropriate constitutive laws that are descriptions of the energy-strain or stress-strain relations of the membrane. It is then possible to model strain-softening membranes, i.e. the membrane stress increases slower
than strain, or strain-hardening membranes, where the membrane stress increases faster
than strain \cite{Dodson_Dimitrakopoulos_2009}. Although constitutive laws are rigorous physical descriptions of the membrane, the numerical implementation is complicated. In addition, due to the continuity of the membrane properties, this approach is limited to modelling at length scales where local differences in the membrane are insignificant \cite{Imai_Omori_Shimogonya_Yamaguchi_Ishikawa_2016}. Another consequence of applying a constitutive model in the entire membrane domain is that the shape memory might be neglected in some cases \cite{Fischer2004}. This approach is discussed in subsection \ref{subsec:continuum}. 

In the second approach, the membrane is treated as a discrete network of springs and particles. This approach provides some advantages namely, its mathematical description is simpler and local differences or thermal fluctuations can be included in contrast to the continuum approach. However, spring constants can depend strongly on the mesh configuration as shown in Ref. \cite{omori_2011}. In fact, Ref. \cite{omori_2011} shows that discrete spring models have anisotropic mechanical properties, in general. Furthermore, when studying biological membranes, local area incompressibility has to be enforced which is not possible using simple discrete spring models \cite{omori_2011}. Additional terms have to be included. This approach is also known to suffer less numerical instability than the continuum approach \cite{Imai_Omori_Shimogonya_Yamaguchi_Ishikawa_2016}. This discrete approach is discussed in subsection \ref{subsec:discrete}. 

The elastic forces are treated differently in each approach. In the continuum approach, these are solved using continuum-based solvers such as the finite element method (FEM) while in the discrete approach, the motion and interaction of individual springs are treated in a particle-based manner. Each approach has its own advantages and drawbacks which we highlight in the following sections. The bending along with the volume and area constraints are in common with both the continuum and discrete approaches and are discussed in section \ref{sec:bending_and_areavolume_constraints}. Additionally, the fluid can be coupled with membrane dynamics using the immersed boundary method (IBM). In subsection \ref{subsec:IBM}, we discuss how to couple the fluid mesh with the membrane mesh. This review focuses on LBM for the fluid which has a Cartesian fixed mesh and for the membrane a (triangular) evolving mesh. The IBM has proven to be useful for partitioned simulations, where the flow and the displacement of the structure are calculated separately \cite{Sun_Bo_2015} and where the elastic bodies have negligible mass \cite{Bottino_1998}. In IBM, the flow is solved on an Eulerian grid while the membrane is treated on a Lagrangian grid \cite{peskin_2002}. The total force acting on the membrane is treated as a source term in the NS equation. This means that artificial forces are spread as body forces around the membrane boundary in order to mimic the effects of the moving membrane. In the LBM, this is included in the $\boldsymbol{g}$ term of Eq. (\ref{eq:NS_equations}), which ensures no slip and no flow penetration boundaries. Because the IBM is often employed for structures with negligible inertia (overdamped regime), the velocity of a membrane node is interpolated from the surrounding fluid while the forces acting on the membrane node are spread in the surrounding fluid nodes. IBM is discussed in more detail in subsection \ref{subsec:IBM}.

\begin{figure}
\center
\includegraphics[width=0.475\textwidth]{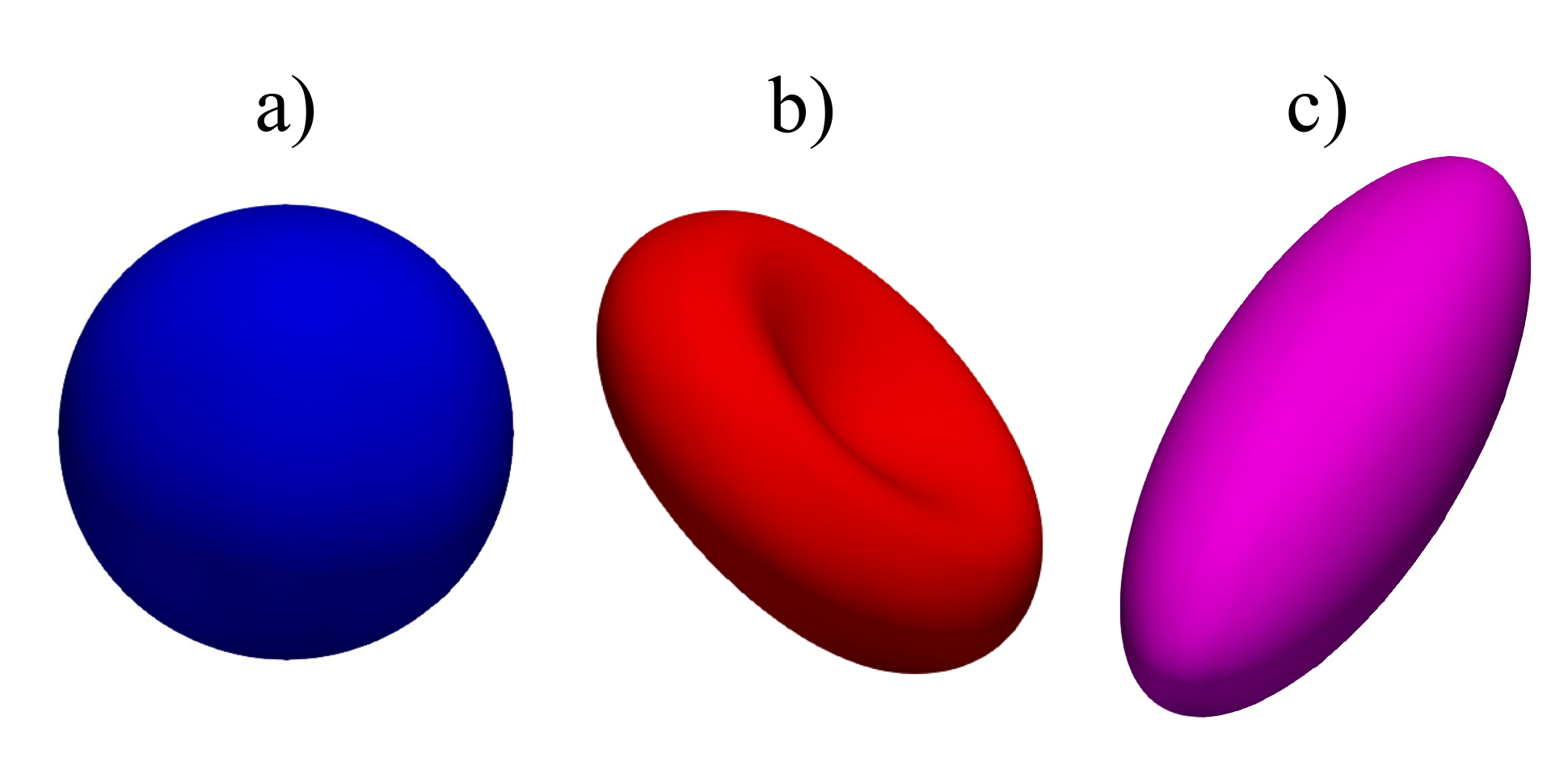}
\caption{ Examples of different shapes that can be used in the lattice-Boltzmann method with finite elements and immerse boundary. (a) is a sphere, (b) is a biconcave shell and (c) is an ellipsoid. An icosahedron was used for the subdivision scheme and there were 5 subdivisions, so the total number of elements is $N_e^5 = 20 \times 4^5 = 20480$ faces.}
\label{fig:membrane_mesh}
\end{figure}

Once we define the constitutive models and constraints, it is straightforward to derive the nodal forces as will be shown. However, it is nontrivial to implement the force calculations. The vast majority of meshes used in the literature are triangular meshes in both the continuum and discrete approaches. We aim, in the following subsections, to describe approaches to obtain those forces for elastic, immersed bodies. We illustrate different possible shapes with the fluid-structure-based models in Fig. \ref{fig:membrane_mesh}.

\subsection{Continuum approach}
\label{subsec:continuum}
The membrane can be treated as a continuum sheet with infinitesimal thickness. The total energy of a membrane is
\begin{equation}
    W = W_s + W_b + W_A + W_V ,
\end{equation}
where $W_s$ is the elastic strain (in-plane) energy,  $W_b$ is the bending energy, $W_A$ is the area penalty energy, and $W_V$ is the volume penalty energy. We highlight that $W_s$ is described by a continuum function which describes the in-plane energy. It is possible then to calculate elastic in-plane forces from $W_s$ using FEM. Moreover, explicit expressions can be used for the calculation of bending and area/volume penalty forces as this does not necessarily require the FEM. In general, to solve the physics of the continuum membrane we need to discretize it using an appropriate mesh (usually triangular). The elementary surfaces in the mesh are connected by nodes which interact with the fluid. The force acting on membrane node $i$ at position $\boldsymbol{x}_i$ is given by the principle of virtual work \cite{charrier_free_1989,shrivastava_large_1993, goldstein2002classical, Lzaro2017},
\begin{equation}
\boldsymbol{F}_i=-\frac{\partial W}{\partial \boldsymbol{x}_i}.
\label{eq:virtual_work_force_continuum}
\end{equation}
Details about derivations using the equation for each term can be found in Ref. \cite{kruger_computer_2012}. For simplicity, we usually consider more straightforward expressions for the force instead of the full expressions derived in the next sections.

\subsubsection{Elastic membrane models and finite element method}
\begin{figure}
\center
\includegraphics[width=0.48\textwidth]{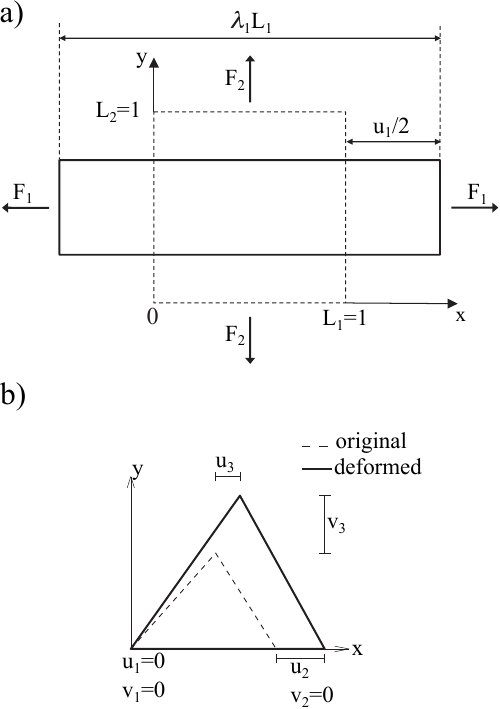}
\caption{a) Example of a deformation of a 2D sheet due to forces $F_1$ and $F_2$. Deformation in the $x$ direction is characterised by the stretch ratio $\lambda_1$ or by the displacement $u_1$ (the same in the $y$ direction). $\lambda_1$ and $\lambda_2$ are the stretch ratios in two principle directions, defined
as the ratio of the final to the initial length. b) Computation of the deformation is done by mapping both the original and the deformed triangles to a common plane.}
\label{fig:principle_strech_ratio}
\end{figure}

To determine the elastic forces acting on membrane nodes, it is essential to choose an appropriate constitutive equation that accurately captures the membrane's inherent properties. We refer the reader to Ref. \cite{barthes-biesel_effect_2002,Lac_Barthes-Biesel_Pelekasis_Tsamopoulos_2004, hashemi2015lattice} for commonly used models which we also discuss in this section. The Neo-Hookean (NH) model \cite{barthes-biesel_effect_2002} is one of the most straightforward choices among different constitutive equations, in the form
\begin{equation}
    W_{s}^{\mathrm{NH}}=\int_A \frac{E_s}{6}\left(\lambda_1^2+\lambda_2^2-3+\frac{1}{\lambda_1^2 \lambda_2^2}\right) \mathrm{d}A , 
    \label{eq:neohookean}
\end{equation}
where $E_s$ is the surface shear elasticity modulus, $\lambda_1$ and $\lambda_2$ are the principle stretch ratios (see Fig. \ref{fig:principle_strech_ratio}a) and $\mathrm{d}A$ is a surface element area. The $\mathrm{NH}$ constitutive equation describes the membrane as a material that doesn't change volume when deformed. If the membrane area increases, it gets thinner to keep the volume constant. An alternative form of Eq. \ref{eq:neohookean} is the Zero-Thickness (ZT) shell representation \cite{ramanujan_deformation_1998} with the strain energy function given as:

\begin{equation}
\begin{aligned}
W_{s}^{\mathrm{ZT}}=& \int_A \frac{E_s}{6}\left[\lambda_1^2+\lambda_2^2-2-\log \left(\lambda_1^2 \lambda_2^2\right)\right.\\
&\left.+\frac{1}{2} \log ^2\left(\lambda_1^2 \lambda_2^2\right)\right] \mathrm{d}A \ \ .
\end{aligned}
\label{eq:zerothickness}
\end{equation}

A widely used model suggested by Skalak \cite{skalak_strain_1973}; see e.g. Ref. \cite{kruger_efficient_2011,sui_dynamic_2008}. The membrane is treated as a thin, 2D elastic surface due to its small thickness compared to its surface area. This allows for an effective modelling of the membrane's behaviour under both low stress and strong deformations. In this model, the strain energy density function is defined as:
\begin{equation}
\begin{aligned}
W_{s}^{\mathrm{SK}}=& =\int_A \frac{E_s}{12}\left[\left(\lambda_1^4+\lambda_2^4-2 \lambda_1^2-2 \lambda_2^2+2\right)\right.\\
&\left.+C\left(\lambda_1^2 \lambda_2^2-1\right)^2\right] \mathrm{d}A,
\end{aligned}
\label{eq:skalak}
\end{equation}
where $E_{s}$ and $C$ are the membrane elastic shear and area dilation moduli. The same is true for other elastic models. Equations (\ref{eq:neohookean}), (\ref{eq:zerothickness}), and (\ref{eq:skalak}) can also be written as a function of the strain invariants $I_{1}$ and $I_{2}$ defined for a 2D membrane (with isotropic and homogeneous properties) as
\begin{equation}
\begin{gathered}
I_{1}=\lambda_{1}^{2}+\lambda_{2}^{2}-2, \\
I_{2}=\lambda_{1}^{2} \lambda_{2}^{2}-1.
\end{gathered}
\end{equation}
We note that some of the methods, such as that proposed by Ref. \cite{Barthes-Biesel_Rallison_1981}, allow one to model a more generic capsule whose interface is also a 2D isotropic sheet with no bending energy. The membrane energy is given by $ W=W\left(I_1, I_2, \alpha_1, \alpha_2, \alpha_3\right)$ with
\begin{align}
W&=W_{0}+\frac{1}{2}\left(\alpha_1-\alpha_3\right) \log \left(I_2+1\right) +\frac{1}{8}\left(\alpha_1+\alpha_2\right) \log ^2\left(I_2+1\right) \nonumber\\
&+\alpha_3\left[\frac{1}{2}\left(I_1+2\right)-1\right],
\end{align}
where $W_0$ is a reference value, $I_1$ and $I_2$ are the strain invariants and $\alpha_1$, $\alpha_2$ and $\alpha_3$  are coefficients which control the elasticity and deformation of the capsule. In fact, one can tune these coefficients to obtain other models such as the Skalak or Neo-Kookean models previously mentioned and even the properties of a droplets \cite{pelusi2023sharp}. 

Once the constitutive law is chosen, we can compute the elastic forces using the linear FEM \cite{charrier_free_1989}. The elastic forces are calculated at the Lagrangian nodes of the membrane mesh, using a linear FEM, commonly employed in capsule deformation studies \cite{Sui_Chew_Low_2007, kruger_efficient_2011,Vahidkhah_Diamond_Bagchi_2014}. These forces result from the planar deformation of discrete surface elements deviating from their original configurations. To enable a comparison between the current and reference configurations, each deformed surface element is projected onto a common plane, aligning with its initial configuration, as shown in Fig. \ref{fig:principle_strech_ratio}b. Triangular surface elements are often used to discretize the membrane thus we also use triangular surface elements in this review. Next, the in-plane displacements, denoted as $u_x$ and $u_y$, can be calculated at every vertex. The procedure, as described below, is motivated by Ref. \cite{charrier_free_1989,shrivastava_large_1993, charrier1987free,armstrong2021numerical,armstrong2021electrohydrodynamic,kruger_efficient_2011,kruger_computer_2012}

Let us denote a undeformed membrane marker described by the reference coordinates $x$ and $y$ and after deformation by $X$ and $Y$. The relation between the undeformed and deformed state is given by:
\begin{equation}
\begin{aligned}
& X=x+u_x, \\
& Y=y+u_y.
\end{aligned}
\label{eq:undeformed_to_deformed}
\end{equation}
As the membrane is modelled using an 2D energy law $W$, the nodal forces, $F_x^{s,i}$ and $F_y^{s,i}$, are derived using the principle of virtual work for a discrete element. This is expressed in terms of virtual displacements, $\delta u_x^i$ and $\delta u_y^i$, given by:
\begin{equation}
\delta W_s=\sum_{i=1}^3\left(\delta u_x^i F_x^{s,i}+\delta u_y^i F_y^{s,i}\right)
\label{eq:virtual_Work_force1}
\end{equation}
where, as in  Fig. \ref{fig:principle_strech_ratio}b, we use the superscripts $i=1,2,3$ to identify the variables vertices of the triangle. $\delta W_s$ is the first order variation in the strain energy of the discrete surface element.

\begin{figure}
\center
\includegraphics[width=0.48\textwidth]{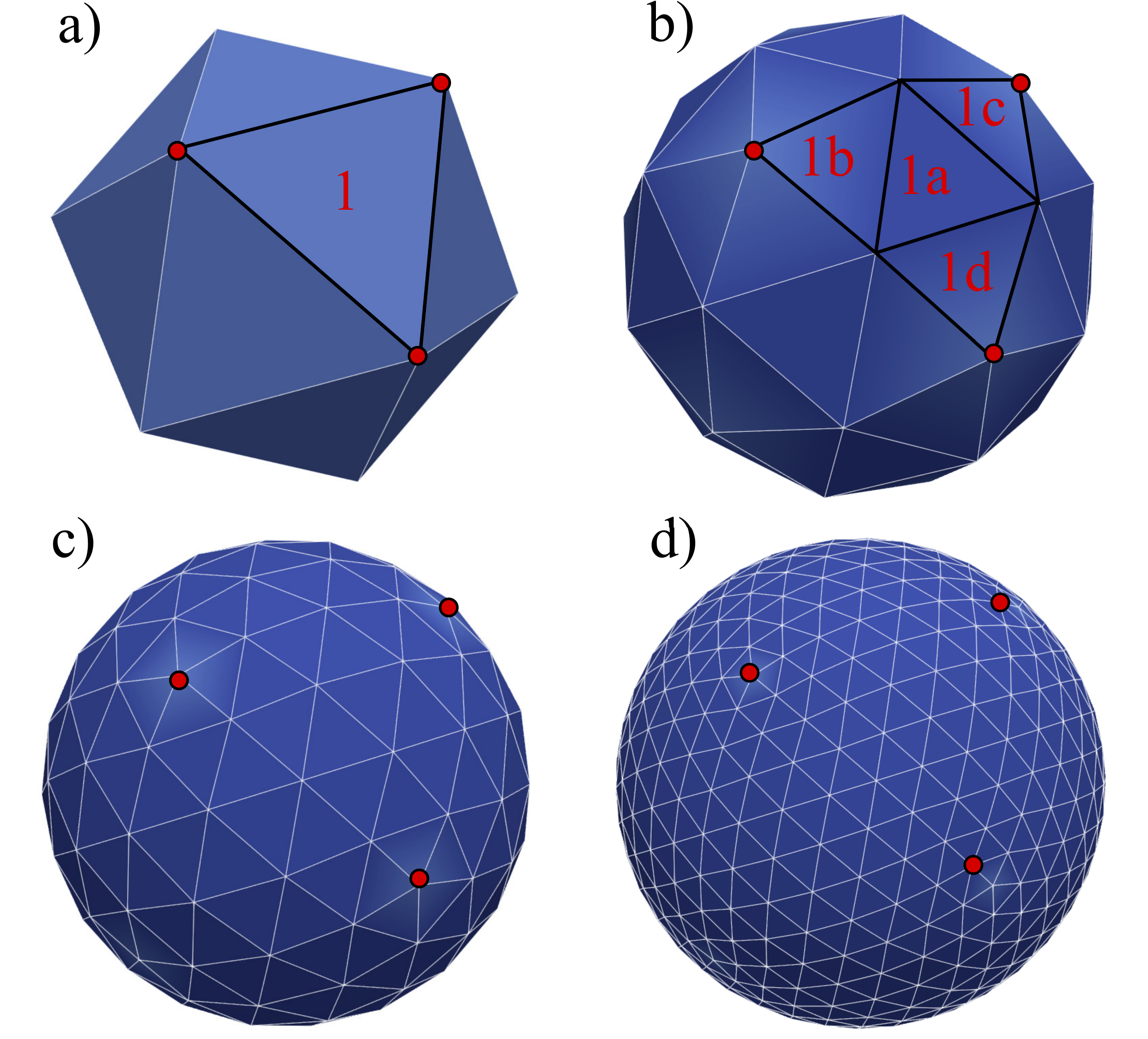}
\caption{Loop subdivision mesh. The mesh is started from a solid such as an icosahedron (or other initial shapes such as a pyramid) which has 20 sides and 12 nodes. A triangular surface element is divided into 4. There are 3 levels of refinement shown: a) to b), b) to c) and c) to d). For clarification, we highlight an initial triangle 1 which is divided into triangles 1a, 1b, 1c, and 1d. The red points just indicate the nodes of the initial face element.}
\label{fig:LoopSbdivisionMesh}
\end{figure}

There are two assumptions one can make for simplicity. Firstly, a discrete element can be assumed to undergo uniform deformation which leads to the stretch ratios being constant on that discrete element. Consequently, the virtual work for an discrete element is given by
\begin{equation}
\delta W_s=A_0 \delta W,
\label{eq:virtual_work_element}
\end{equation}
where $A_0$ is the original area of the discrete element and $W$ strain energy density. Secondly, one can assume that the strain energy is related to the principal stretch ratios, $\lambda_1$ and $\lambda_2$ due to the membrane being isotropic and incompressible. This leads to the first variation of the strain energy function, $\delta W$, computed using the chain rule resulting in the following equation for virtual work of a discrete element due to displacements, $\delta u_x^i$ and $\delta u_y^i$:
\begin{multline}
\delta W_s=A_0 \sum_{k=1}^3 \left[ \delta u_x^k\left(\frac{\partial \lambda_1}{\partial u_x^k} \frac{\partial W}{\partial \lambda_1}+ \frac{\partial \lambda_2}{\partial u_x^k} \frac{\partial W}{\partial \lambda_2}\right)+ \right. \\ \left. \delta u_y^k \left(\frac{\partial \lambda_1}{\partial u_y^k} \frac{\partial W}{\partial \lambda_1}+\frac{\partial \lambda_2}{\partial u_y^k} \frac{\partial W}{\partial \lambda_2} \right) \right] \ \ .
\label{eq:virtual_Work_force2}
\end{multline}
To obtain expressions for the nodal forces, $F_x^{s,i}$ and $F_y^{s,i}$, Eqs. (\ref{eq:virtual_Work_force1}) and (\ref{eq:virtual_Work_force2}) are substituted in Eq.(\ref{eq:virtual_work_element}) and using the fact that displacements, $\delta u_x^i$ and $\delta u_y^i$, are arbitrary gives:
\begin{equation}
\begin{aligned}
F_x^{s,i} & =A_0 \frac{\partial W}{\partial \lambda_1} \frac{\partial \lambda_1}{\partial u_x^i}+A_0 \frac{\partial W}{\partial \lambda_2} \frac{\partial \lambda_2}{\partial u_x^i}, \quad i=1,2,3 \\
F_y^{s,i} & =A_0 \frac{\partial W}{\partial \lambda_1} \frac{\partial \lambda_1}{\partial u_y^i}+A_0 \frac{\partial W}{\partial \lambda_2} \frac{\partial \lambda_2}{\partial u_y^i}, \quad i=1,2,3 ,
\end{aligned}
\label{eq:virtual_work_final}
\end{equation}
The forces $F_x^{s,i}$ and $F_y^{s,i}$ are in the local coordinate system, which after computing are then transformed back to the global Cartesian coordinates and added to the node's Lagrangian force term as shown in Ref. \cite{charrier_free_1989}. As an example, in a triangular mesh, every node serves as the vertex for 5 or 6 triangles as shown in Fig. \ref{fig:LoopSbdivisionMesh} where we highlight a few Lagrangian nodes in red. Consequently, the total elastic force at each node is the combined force from each of these 5 or 6 triangles.

The derivatives in Eq. (\ref{eq:virtual_work_final}) are computed using a linear FEM, with linear shape functions, $N^i$, $i=1,2,3$, where  $N^1$, is defined as \cite{charrier_free_1989,armstrong2021electrohydrodynamic,armstrong2021numerical}
\begin{align}
N^1(x, y)&=\frac{1}{2 A_0}\left[\left(y^{(2)}-y^{(3)} k\right) x+\left(x^{(3)}-x^{(2)}\right) y \right.\nonumber \\ &\left. +x^{(2)} y^{(3)} k-x^{(3)} y^{(2)}\right]
\end{align}
and $N^2$ and $N^3$ can be obtained by cycling the indices from $1 \rightarrow 2 \rightarrow 3 \rightarrow 1$. $2 A_0$ is twice the area of the undeformed element and is computed as follows:
\begin{align}
2 A_0&=\left|\left(y^{(2)}-y^{(3)}\right) x^{(1)}+\left(x^{(3)}-x^{(2)}\right) y^{(1)}+x^{(2)} y^{(3)} \right.\nonumber \\ &\left. -x^{(3)} y^{(2)}\right| \ \ .
\end{align}
In turn, the spatial dependence of the displacement on the element can be expressed as
\begin{equation}
\begin{aligned}
& u_x(x, y)=N^1(x, y) u_x^{(1)}+N^2(x, y) u_x^{(2)}+N^3(x, y) u_x^{(3)} \\
& u_y(x, y)=N^1(x, y) u_y^{(1)}+N^2(x, y) u_y^{(2)}+N^3(x, y) u_y^{(3)} \ \ .
\end{aligned}
\label{eq:spacial_dependence}
\end{equation}
The relation between $(x, y)$ and $(X, Y)$  (see Eq. (\ref{eq:undeformed_to_deformed})) is given in matrix notation by $\mathbf{d X}=\mathbf{F} \cdot \mathbf{dx}$ where the matrix of the Cartesian components of the deformation gradient tensor is $\mathbf{F}=\mathbf{\frac{\partial X}{\partial x}}=\mathbf{I}+\mathbf{\frac{\partial u}{\partial x}}$ as shown in Ref. \cite{charrier_free_1989,shrivastava_large_1993}.
Lastly, one just needs to compute the principle stretch ratios, $\lambda_1$ and $\lambda_2$, from the eigenvalues of the two-dimensional right Cauchy-Green tensor, $\mathbf{G}=\mathbf{F}^\mathbf{T} \mathbf{F}$, whose components are given by

\begin{equation}
\begin{aligned}
G_{11} & =\left(1+\frac{\partial u_x}{\partial x}\right)^2+\left(\frac{\partial u_y}{\partial x}\right)^2 \\
G_{22} & =\left(1+\frac{\partial u_y}{\partial y}\right)^2+\left(\frac{\partial u_x}{\partial y}\right)^2 \\
G_{21}=G_{12} & =\left(1+\frac{\partial u_x}{\partial x}\right)\left(\frac{\partial u_y}{\partial y}\right)+\left(1+\frac{\partial u_y}{\partial x}\right)\left(\frac{\partial u_x}{\partial y}\right) .
\end{aligned}
\label{eq:Chauchy_green}
\end{equation}
The derivatives of the in-plane displacements, $u$ and $v$, in Eq. (\ref{eq:Chauchy_green}) for $G_{21}=G_{12}$ are computed by differentiating their linear representations given by Eq. (\ref{eq:spacial_dependence}). Finally, the principle stretch ratios are expressed in terms of the components of $\boldsymbol{G}$ as follows:
\begin{equation}
\begin{aligned}
& \lambda_1^2=\frac{1}{2}\left[G_{11}+G_{22}+\sqrt{\left(G_{11}-G_{22}\right)^2+4 G_{12}^2}\right] \\
& \lambda_1^2=\frac{1}{2}\left[G_{11}+G_{22}-\sqrt{\left(G_{11}-G_{22}\right)^2+4 G_{12}^2}\right] .
\end{aligned}
\end{equation}


\subsection{Discrete approach}
\label{subsec:discrete}
A different approach to model membrane elasticity is based on molecular dynamics, as initially suggested in Ref. \cite{Li_2007}, and subsequently widely used \cite{FEDOSOV_2010,TSUBOTA_2006}. Although theoretically possible to use molecular dynamics to model fluid-structure interaction flows, the computational cost of molecular dynamics is prohibitively expensive to reach the length and time scales relevant to the motion and deformation of flexible objects. Addressing this regime requires a mesoscale method. One strategy involves using a discrete arrangement of springs and (coarse-grained) particles. As one represents multiple atoms as a single particle, the computational cost is reduced. In addition, this simplification helps in studying larger spatial scales or longer timeframes while sacrificing some fine-scale details. Furthermore, this allows for defining an effective interaction potential between each particle to simulate the mechanical behaviour of the system. By adopting this approach, the number of degrees of freedom is minimised, and the time and length scales are increased. The force acting in each particle can be obtained by taking the derivative of the potential energy. Other forces such as fluid forces and particle-particle repulsion are also taken into account. 

While membranes have different constituents, most share common physical properties. The in-plane stretching energy can be modelled through several different potentials. These can range from the simple linear such as the harmonic potential to a complex nonlinear ones such as worm-like-chain (WLC) potential or the finite extensible nonlinear elastic (FENE) potential. This approach has proven effective in making predictions that consistently match the energy density functions used in the continuum model \cite{Li_2007,omori_2011}. The total energy of a membrane is
\begin{equation}
W =W_{\text {stretch }}+W_{\text {bending }}+W_{\text {area }}+W_{\text {volume }},
\end{equation}
where $W_{\text{volume}}$ is included only for 3D simulations. Once again, the nodal force at node $i$ at position $\boldsymbol{x}_{i}$ derived from the potential energy is given by
\begin{equation}
\boldsymbol{F}_{i}=-\frac{\partial W\left(\boldsymbol{x}_{i}\right)}{\partial \boldsymbol{x}_{i}}.
\end{equation}
Full derivations are not shown but can be found in Ref.~\cite{pozrikidis_computational_2010}. 

\subsubsection{Elastic membrane models}

A harmonic stretching potential can be used to model the stretching energy as
\begin{equation}
W_{\text {stretch }}= W_{\text {HAR}}=\frac{1}{2} k_{s} \sum_{j=1, \ldots, N_{s}}\left(l_{j}-l_{j 0}\right)^{2},
\label{eq:harmonic_potential}
\end{equation}
where $k_{s}$ is the linear spring stretching constant, $N_S$ is the number of springs, $l_{j}$ and $l_{j 0}$ are length and the equilibrium length of the $j$-th spring, respectively. Eq. (\ref{eq:harmonic_potential}) has been used  to study diseased $\mathrm{RBCs}$ \cite{wu_simulation_2013}, blood flow in capillaries \cite{nakamura_spring-network-based_2013} and has been validated against modern experimental techniques such as atomic
force microscopy \cite{barns2017investigation}. Eq. (\ref{eq:harmonic_potential}) is one of the simplest form of in-plane to model in-plane streching energy, however it is restricted in its ability to model nonlinear characteristics. Nonetheless, some attempts have been made to account for an increase in stretching resistance with elongation of the spring by using a nonlinear spring constant $k_{s}$ as $k_{s}=k_{s 0} e^{2(\iota-1)}$ where the bond stretch ratio $(\iota)$ is included \cite{nakamura_analysis_2014}. The stretch ratio $\iota$ is defined as $\iota=l / l_{0}$ where $l$ and $l_{ 0}$ are length and the equilibrium length of the spring, respectively. It has been shown to achieve good results between experiments and simulations \cite{nakamura_spring-network-based_2013}. Similarly, other nonlinear potentials may be used to model the stretching energy. The two most popular nonlinear potentials \cite{FEDOSOV_2010} are the WLC potential and FENE potential. Explicitly, they are defined as
\begin{equation}
W_{\text{WLC}}=\frac{k_{B} T l_{m}}{4 p} \frac{3 x^{2}-2 x^{3}}{1-x}, \quad \quad W_{\text{FENE}}=-\frac{k_{s}}{2} l_{m}^{2} \log \left[1-x^{2}\right],
\label{eq:WLC_FENE}
\end{equation}
where $k_{B}$ is the Boltzmann constant, $p$ is the persistence length, $T$ is the temperature, $x=l / l_{m} \in(0,1), l_{m}$ is the maximum spring extension and $k_{s}$ is the FENE spring constant. Notice that in both models the maximum extension is limited to $l_m$ as the force will approach infinity as the spring length approaches $l_m$.

Note that these springs (FENE and WLC) represent attractive potentials, thus they tend to reduce the area (area compression). A repulsive force field should be combined to restrict this reduction. Ref. \cite{FEDOSOV_2010} proposed two potentials, one of which is an inverse power repulsive potential ($W_{\text{POW}}$) that is based on the spring length i.e it will restrict the length of the spring. This means
\begin{equation}
W_{\text{stretch}} = W_{\text{WLC}} +  W_{\text{POW}} \quad \quad \text{or} \quad \quad W_{\text{stretch}} = W_{\text{FENE}} +  W_{\text{POW}},
\end{equation}
where the inverse power repulsive potential is given by
\begin{equation}
W_{\text{POW}}=\frac{k_{p}}{(n-1) l^{n-1}} \quad n>0, n \neq 1
\label{eq:inversePower}
\end{equation}
and $k_p$ is the repulsive stiffness. Finally, the forces for Eqs. (\ref{eq:WLC_FENE}) and (\ref{eq:inversePower}) are
\begin{equation}
\begin{aligned}
&\boldsymbol{F}_{\text{WLC}}(l)=-\frac{k_B T}{p}\left(\frac{1}{4(1-x)^2}-\frac{1}{4}+x\right) \hat{\boldsymbol{l}}_{i j}, \\
&\boldsymbol{F}_{\text{FENE}}(l)=-\frac{k_s l}{1-x^2} \hat{\boldsymbol{l}}_{i j}, \\
&\boldsymbol{F}_{\text{POW}}(l)=\frac{k_p}{l^m} \hat{\boldsymbol{l}}_{i j},
\end{aligned}
\end{equation}
where $x=l / l_m$ and $\hat{\boldsymbol{l}}_{i j}=\vec{l}_{i j} / l$ is the vector of unit length between nodes $i$ and $j$.

\subsection{Bending and area \& volume constraints}
\label{sec:bending_and_areavolume_constraints}
\subsubsection{Bending}
The resistance to bending and the prevention of buckling is relevant in many membrane-bounded bodies, particularly biological ones. The implementation of bending is not trivial and it is computationally intricate. In fact, it has been the subject of several reviews \cite{guckenberger_bending_2016,guckenberger_theory_2017,bian2020bending}. Since there are many ways to implement bending, for simplicity, we focus on the two most common approaches, one for the continuum approach and another for the discrete approach. We refer the reader to Ref.   \cite{guckenberger_bending_2016,guckenberger_theory_2017} for bending models and numerical implementations. In line with the continuum approach, one of the most popular bending formulations is that of Helfrich \cite{helfrich_elastic_1973}
\begin{equation}
W_b=\frac{E_B}{2} \int_A\left(2 \kappa-c_0\right)^2 d A+E_G \int_S \kappa_G \mathrm{d}A,
\end{equation}
where  $ \kappa$ is the mean curvature and  $ \kappa_G$ is the Gaussian curvature, $c_0$ is the spontaneous curvature (curvature for which the bending energy is minimal). It has been shown that the bending force acting on node $i$ can be written as \cite{zhong-can_bending_1989,sinha_dynamics_2015}
\begin{equation}
\boldsymbol{F}_{i}^{b}=E_B\left[\left(2 \kappa+c_0\right)\left(2 \kappa^2-2 \kappa_g-c_0 \kappa\right)+2 \Delta_{\mathrm{LB}} \kappa\right] \boldsymbol{n}
\label{eq:bending_simple}
\end{equation}
where $\boldsymbol{n}$ is the outwards normal vector to the surface and $\Delta_{LB}$ is the Laplace-Beltrami operator. Calculation of Eq. (\ref{eq:bending_simple}) and $ \Delta_{\mathrm{LB}} \kappa$ can be found in Ref. \cite{guckenberger_bending_2016,guckenberger_theory_2017,reuter_discrete_2009} .

Another popular and simplified expression for the bending energy \cite{bian2020bending} often used with the discrete approach is,
\begin{equation}
W_{\text {bending }}= k_{b} \sum_{j \in 1 \ldots N_{b}} \left[1 -\cos\left(\theta_{k}-\theta_{k 0}\right)  \right],
\label{eq:bending_spring}
\end{equation}
where $k_{b}$ is the spring constant for bending, $N_b$ is the number of bending springs, $\theta_{k}$ is the angle between two outward surface normals of two neighbouring triangular elements that share the edge $k$ and $\theta_{k0}$ is the equilibrium angle. The corresponding force Eq. (\ref{eq:bending_spring}) is given by
\begin{equation}
\boldsymbol{F}_{i}^b=-k_b \sin \left(\theta_k-\theta_{k0}\right) \frac{\partial \theta_{k}}{\partial \boldsymbol{x}_i}.
\end{equation}
Details on the calculation of $\frac{\partial \theta_{k}}{\partial \boldsymbol{x}_i}$ can be found in Ref. \cite{kruger_computer_2012,bian2020bending}.

\subsubsection{Area and volume constraints}
In order to impose area and volume conservation it may be necessary to include appropriate constraints. This will depend on the problem at hand. In some cases like a capsule in shear flow, the volume change is negligible and it is not necessary to impose volume conservation. The area and volume energy constraints can be enforced by adding the  energy terms \cite{kruger_computer_2012}
\begin{equation}
\begin{aligned}
W_A &=\frac{k_a}{2} \frac{\left(A-A_0\right)^2}{A_0}, \\
W_V &=\frac{k_v}{2} \frac{\left(V-V_0\right)^2}{V_0} .
\end{aligned}
\label{eq:contraints_continuum}
\end{equation}
where $k_a$ and $k_v$ are the area and volume coefficients that control the strength of the force which regulates the change in total area $A$ and the volume $V$ from the initial area $A_0$ and volume $V_0$. While $A_0$ and $V_0$ are input parameters, $A$ and $V$ need to be computed at every time step. From Eqs. (\ref{eq:contraints_continuum}) we find that the corresponding forces are
\begin{equation}
\begin{aligned}
\boldsymbol{F}_{i}^{A} &=k_a \frac{\left(A-A_0\right)}{A_0}\frac{\partial A}{\partial \boldsymbol{x}_i}, \\
\boldsymbol{F}_{i}^{V} &=k_v \frac{\left(V-V_0\right)}{V_0}\frac{\partial V}{\partial \boldsymbol{x}_i}.
\end{aligned}
\label{eq:contraints_continuum_force}
\end{equation}
More details and computation of $\frac{\partial A}{\partial \boldsymbol{x}_i}$ and $\frac{\partial V}{\partial \boldsymbol{x}_i}$ may be found in Ref. \cite{kruger_computer_2012}. While $A$ can be calculated by summing the elemental areas (e.g. triangles), the calculation of volume requires a more sophisticated approach. An efficient way to calculate $V$ is described in Ref. \cite{cha_zhang_efficient_2001}. We can also express Eqs. (\ref{eq:contraints_continuum}) as \cite{gou_effects_2019},
\begin{equation}
\begin{aligned}
\boldsymbol{F}_{i}^A &= -k_a \frac{A-A_0}{A_0} \boldsymbol{n}, \\
\boldsymbol{F}_{i}^V &= -k_v \frac{V-V_0}{V_0} \boldsymbol{n},
\end{aligned}
\label{eq:contraints_continuum_force_simpler}
\end{equation}
where $\boldsymbol{n}$ is the outwards normal to the surface. These constraints for area and volume are also used in models which employ a discrete approach.

Finally,  since a node can be the vertex of multiple surface elements, the total force $\boldsymbol{F}_{i} = \boldsymbol{F}_{i}^{s} +  \boldsymbol{F}_{i}^{b} +  \boldsymbol{F}_{i}^{A} +  \boldsymbol{F}_{i}^{V}$ acting on node $i$ is, 
\begin{equation}
\boldsymbol{F}_{i}=\sum_j\boldsymbol{F}_{i,j},
\end{equation}
where $j$ runs over each surrounding element.

\subsection{Linking fluid to structure}
\label{subsec:IBM}

The field of fluid-structure interaction (FSI) revolves around the intricate dynamics and interdependence between two fundamental components: a fluid and a solid. This dynamic interplay gives rise to a range of challenges and opportunities, offering insights into complex phenomena across various scientific and engineering domains, encompassing areas like paint pigments, polymers, gels, proteins, red blood cells flow in the heart, and more {\cite{hirt_arbitrary_1974, tian2014fluid, duprat2016fluid}. 
The essence of FSI lies in understanding how fluids and solids interact dynamically. When a fluid flows around or interacts with a solid structure, it induces forces and deformations in the solid, while the solid, in turn, influences the behaviour of the fluid. This mutual influence creates a complex coupling that is prevalent in numerous natural and engineered systems. However, they pose challenges due to their nonlinear and multiscale nature. The nonlinear interactions between fluid and solid components, coupled with the need to consider multiple scales of phenomena, make analytical solutions often impossible. leading to the development of numerical methods to simultaneously address fluid and structure dynamics. In these methods, interface boundary conditions are of crucial as they ensure consistency and continuity between the fluid and solid domains. They dictate how forces, velocities, and deformations are transmitted across the interface. When considering interfacial conditions, there are two main approaches. The first is the Arbitrary Lagrangian-Eulerian (ALE) method \cite{hirt_arbitrary_1974}. In the ALE method, the mesh used to discretize the fluid domain is allowed to move. Unlike the traditional Eulerian approach where the mesh is fixed in space or the Lagrangian approach where the mesh moves with the fluid, the ALE method allows the motion of the mesh. On the interface between fluid and solid, the velocity and stress should be continuous. The fluid mesh follows the shape of the solid mesh, making it easy to set accurate and efficient boundary conditions at the fluid-solid interface. Yet, when solids undergo significant deformations, re-meshing is often needed to prevent mesh tangling which can affect the mesh motion. This process of mesh regeneration can be challenging during simulations. As an alternative the (IBM) was proposed \cite{peskin_2002} using a non-boundary fitting method. The IBM uses two independent meshes for fluid and solid components. The solid boundaries are immersed in the fluid mesh, and forces from the solid are spread into the fluid. This eliminates the need for the fluid mesh to conform to the solid boundaries.

The mechanics of the solid and fluid flow need to be solved. This can be done in two ways: a partitioned way where the structural mechanics and flow equations are solved separately or a monolithic way where they are solved simultaneously. FSI problems using the LBM often use the IBM as a coupling algorithm. This allows software modularity and different, possibly more efficient algorithms can be used to solve the structural mechanics.
 
This section will discuss the technical details of IBM and its coupling with the LBM and is primarily driven by Ref. \cite{kruger_lattice_2017,tan2015lattice,peskin_three-dimensional_1989,peskin_2002}. Other coupling techniques such as the stress integration approach \cite{kollmannsberger_fixed-grid_2009} and friction coupling approach \cite{ahlrichs_lattice-boltzmann_1998,ahlrichs_simulation_1999,hsu_migration_2010,Dunweg_lattice_2009} are briefly mentioned. The IBM was selected due to its efficiency and popularity. It was first proposed by Peskin to study blood flow in the heart \cite{peskin1972flow,peskin_2002} using vortex methods. The fluid is solved on a stationary mesh, while the immersed solids are modelled using a flexible moving mesh. This mesh is not tied to the fixed structure of the fluid mesh. Information is exchanged between the fluid and solid mesh through nodal interpolation. Two-way coupling implies that both the fluid and the immersed structures influence each other. The motion and forces of the fluid impact the immersed structures, and, conversely, the motion and forces of the immersed structures affect the fluid through an effect force density and can prevent fluid penetration in the solid. IBM has been used in the simulation of jellyfish \cite{herschlag_reynolds_2011}, blood flow \cite{kruger_efficient_2011,zhang_red_2008,liu_rheology_2006} and platelet migration \cite{crowl_computational_2010}. We refer the reader to Ref. \cite{peskin_2002,mittal_immersed_2005} for a more detail description of IBM and applications. 

Firstly, one can start a description of IBM through a Lagrangian perspective. The immersed structure is characterised as a parametric surface, denoted as $X(p, q, r, t)$. This means that the shape and position of the structure are defined by curvilinear coordinates $(p, q, r)$ varying in a 3D space, with the additional temporal parameter $t$ representing time. Curvilinear coordinates are particularly useful for describing complex shapes or deformations that may not align with a Cartesian coordinate system (e.g. fluid domain). In contrast, The fluid domain is described by coordinates which are fixed in space and provide a frame of reference independent of the motion of the fluid i.e. Eulerian coordinates denoted by $\boldsymbol{x}$. Ref. \cite{peskin_three-dimensional_1989,peskin_2002} defines the force density $\boldsymbol{g}(\boldsymbol{x}, t)$ exerted by the structure on the fluid as
\begin{equation}
\boldsymbol{g}(\boldsymbol{x}, t)=\int \boldsymbol{F}(p, q, r, t) \delta(\boldsymbol{x}-\boldsymbol{X}(p, q, r, t)) \mathrm{d}p\, \mathrm{d}q\, \mathrm{d}r.
\label{eqn:force_distribution}
\end{equation}
This force is distributed then used by the momentum equation of the surrounding fluid nodes. $\delta(\boldsymbol{x})$ is the delta function $\delta\left(x\right) \delta\left(y\right) \delta\left(z\right)$ where $x, y, z$ are the Cartesian components of the position vector $\boldsymbol{x}$. Likewise, the structural velocity $\boldsymbol{v}(\boldsymbol{X}(p, q, r, t), t)$ is updated by interpolating the neighbouring fluid velocities $\boldsymbol{u}(\boldsymbol{x}, t)$ as
\begin{equation}
\boldsymbol{v}(\boldsymbol{X}(p, q, r, t), t)=\int \boldsymbol{u}(\boldsymbol{x}, t) \delta(\boldsymbol{x}-\boldsymbol{X}(p, q, r, t)) \mathrm{d}\boldsymbol{x}.
\label{eqn:velocity_interpolation}
\end{equation}
Note that Eq. (\ref{eqn:velocity_interpolation}) represents the continuity of velocity on the fluid-solid boundary.

The idea of IBM coupling is shown in Fig. \ref{fig:two_way_coupling} where the black dots represent fluid nodes, and magenta dots represent the solid structure. How many nodes are used for the interpolation and their weights depend on $\delta(\boldsymbol{x})$. For instance, Fig. \ref{fig:two_way_coupling}(a) illustrates that the velocity of the magenta node $X$ will be interpolated from the surrounding four black fluid nodes within the shaded square frame. Likewise, in Fig. \ref{fig:two_way_coupling}(b) the structural force will spread to the four surrounding nodes.

\begin{figure}[!htbp]
\center
\includegraphics[width=0.49\textwidth]{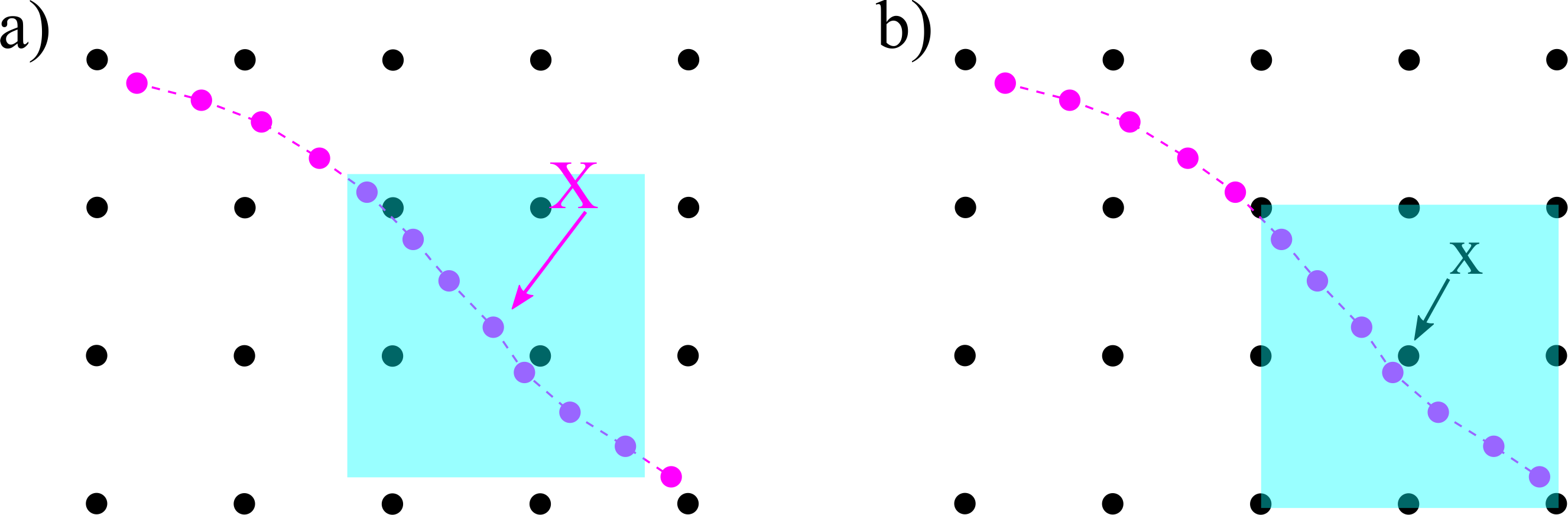}
\caption{IBM two-way fluid-solid coupling. a) The solid velocity $\boldsymbol{v}(\boldsymbol{X}, t)$ is interpolated from neighbouring fluid nodes within the shaded square box. The contribution from a fluid node is weighted by the $\delta(\boldsymbol{x})$ function. b) The solid force $\boldsymbol{g}(\boldsymbol{x}, t)$ will spread to the local fluid nodes as a force density. We see that the fluid node $\boldsymbol{x}$ will receive a force contribution from the magenta nodes within the shaded area as the magenta nodes will spread the force to the nearest four black nodes. As $x$ is the nearest fluid node to the magenta nodes in the shaded area, it will receive the force from them. The spread of the force among fluid nodes is given by the $\delta(\boldsymbol{x})$ function. The number of neighbouring nodes to spread the force can vary.
}
\label{fig:two_way_coupling}
\end{figure}

\subsubsection{Spatial and temporal discretization}
Implementing the IBM requires both spatial and temporal discretization of Eqs. (\ref{eqn:force_distribution}) and (\ref{eqn:velocity_interpolation}), is needed for the numerical implementation of the IBM. Following Ref. \cite{peskin_2002,tan2015lattice}, we address the discretization in space first, then proceed to discuss the selection of $\delta(\boldsymbol{x})$, and, finally the discretization in time. 

The discretization in space of the Eulerian grid, denoted by $g_{x}$ is a uniform and regular grid in $x, y, z$ coordinates, e.g., $\boldsymbol{x}=\left(x_{j}, y_{j}, z_{j}\right) \Delta x$, where $\Delta x$ is the spacing, $\left(x_{j}, y_{j}, z_{j}\right)$ are the components of the position in each direction. This discretization also matches the discretization used in the LBM, which leads to $\Delta x=\delta x$ and  $\Delta t=\delta t$. Equally, the discretization of the Lagrangian grid, denoted by $G_{s}$, is the set of $(p, q, r)$ of the form $\left(p_{k} \delta p, q_{k} \delta q, r_{k} \delta r\right)$, where $\left(p_{k}, q_{k}, r_{k}\right)$ are integers. Ref. \cite{peskin_2002} recommended that $\delta s<\frac{\Delta x}{2}, s \in\{p, q, r\}$ to prevent fluid penetration/leaking. However, Ref. \cite{kruger_computer_2012} systematically studied mesh ratio between solid and fluid for small deformations, e.g., $\frac{\delta s}{\Delta x}$, and concluded that a range of approximately $(0.5,1.5)$ is enough to prevent fluid penetration/leading without notably affecting the results. Nonetheless, if large deformations occurs that the mesh should be refined accordingly to capture these deformations. Resuming, the force spreading equation (\ref{eqn:force_distribution}) becomes
\begin{equation}
\boldsymbol{g}(\boldsymbol{x}, t)=\sum_{(p, q, r) \in G_{s}} \boldsymbol{F}(p, q, r, t) \delta_{\Delta}(\boldsymbol{x}-\boldsymbol{X}(p, q, r, t)) \Delta p\, \Delta q\, \Delta r,
\label{eqn:force_spreading_2}
\end{equation}
where $\boldsymbol{F}(p, q, r, t)$ is the force density from the solid structures. As in Ref. \cite{tan2015lattice}, let us define $\mathcal{F}=\boldsymbol{F}(p, q, r, t) \Delta p\, \Delta q\, \Delta r$. $\mathcal{F}$ can be viewed as the integration of the force density $\boldsymbol{F}$ over the element of volume $\mathrm{d} v=\Delta p\, \Delta q\, \Delta r$, which is the force term applied to each node. The Lagrangian nodes of the solid can be indexed as $i$ without losing generality. Equation (\ref{eqn:force_spreading_2}) is then simplified to
\begin{equation}
\boldsymbol{g}(\boldsymbol{x}, t)=\sum_{i \in G_{s}} \mathcal{F}_{i} \delta_{\Delta}\left(\boldsymbol{x}-\boldsymbol{X}_{i}\right),
\label{eqn:force_spreading_3}
\end{equation}
where $\boldsymbol{X}_{i}$ is the position of the $i$-th Lagrangian node. Eq. (\ref{eqn:velocity_interpolation}) is simplified to
\begin{equation}
\boldsymbol{v}\left(\boldsymbol{X}_{i}, t\right)=\sum_{x \in g_{x}} \boldsymbol{u}(\boldsymbol{x}, t) \delta_{\Delta}\left(\boldsymbol{x}-\boldsymbol{X}_{i}\right) \Delta x^{3}.
\label{eqn:velocity_interpolation_2}
\end{equation}
In the LBM, the time and spatial step in LB units are usually assumed unity, e.g., $\Delta t=1, \Delta x=1$, respectively. Thus, Eq. (\ref{eqn:velocity_interpolation_2}) is simplified to
\begin{equation}
\boldsymbol{v}\left(\boldsymbol{X}_{i}, t\right)=\sum_{x \in g_{x}} \boldsymbol{u}(\boldsymbol{x}, t) \delta_{\Delta}\left(\boldsymbol{x}-\boldsymbol{X}_{i}\right).
\end{equation}
Ref. \cite{peskin_2002} demonstrated that the $\delta_{\Delta}(\boldsymbol{x})$ function must adhere to specific constraints and properties to ensure consistent calculation of mass, force, and torque from both Eulerian and Lagrangian point of view. Here, we assumed that $\delta_{\Delta}(\boldsymbol{x})$ expressed through the scalar function $\phi(x)$
\begin{equation}
\delta_{\Delta}(\boldsymbol{x})=\phi(x) \phi(y) \phi(z),
\end{equation}
where $(x, y, z)$ are the three components of the position vector $\boldsymbol{x}$. We do not delve into the specifics of the $\phi(x)$ function. Rather, our intention is to highlight the frequently utilised four-point interpolation function, which is,
\begin{equation}
\phi(x)= \begin{cases}0, \quad|x| \geq 2 \\ \frac{1}{8}\left(5-2|x|-\sqrt{-7+12|x|-4 x^{2}}\right), & 1 \leq|x| \leq 2 \\ \frac{1}{8}\left(3-2|x|+\sqrt{1+4|x|-4 x^{2}}\right), & 0 \leq|x| < 1 .\end{cases}
\label{eqn:four_point_interpolation_function}
\end{equation}
Ref. \cite{peskin_2002} showed that $\phi(x)$ can be accurately estimated using a straightforward formula
\begin{equation}
\phi(x)= \begin{cases}0, \quad \text { otherwise } \\ \frac{1}{4}\left(1+\cos \left(\frac{\pi x}{2}\right)\right), & |x| \leq 2\end{cases}.
\end{equation}
However, Eq. (\ref{eqn:four_point_interpolation_function}) is much faster as it involves evaluating polynomial functions rather than the $\cos (x)$ function. Given that the delta function is used to transfer quantities between the solid and the fluid, the choice of its regularisation determines the accuracy of the IBM. Certain properties have to be satisfied in a discrete way by the discrete version of the delta function, to ensure the conservation of force, mass and torque, or to prevent void jumps of variables on the grid nodes \cite{kajishima2016computational}. Often three or four-grid approximations are sufficient to satisfy the necessary discrete conditions. Linear interpolation is too simplistic of an approximation and does not lead to the conservation of the physical quantities. Studies of the stability of different regularised delta functions and smoothing techniques can be found in Ref. \cite{kajishima2016computational,shin2008assessment, yang2009smoothing}.

Because the fluid and solid are solved in an alternating fashion, the IBM follows a partitioned approach. To achieve a temporal discretization scheme with second-order accuracy, Ref. \cite{peskin_three-dimensional_1989,peskin_2002} suggested an integration scheme based midpoint rule \cite{peskin_2002}. For the sake of simplicity, the solution at time step $n$ will be denoted by a superscript on the variable. We can then consider the solid nodal position $X_{i}^{n}$ at time step $n$, the intermediate position at time step $n+\frac{1}{2}$ is calculated using
\begin{equation}
\boldsymbol{X}_{i}^{n+\frac{1}{2}}=\boldsymbol{X}_{i}^{n}+\frac{1}{2} \sum_{\boldsymbol{x} \in g_{x}} \boldsymbol{u}^{n} \delta_{\Delta}\left(\boldsymbol{x}-\boldsymbol{X}_{i}\right) .
\label{eqn:intermidiate_position}
\end{equation}
Using the updated solid nodal position at the $n+\frac{1}{2}$ time step, the force acting on the structure can be evaluated as
\begin{equation}
\mathcal{F}_{i}^{n+\frac{1}{2}}=\boldsymbol{F}_{i}^{n+\frac{1}{2}} \Delta p\, \Delta q\, \Delta r=-\frac{\partial}{\partial \boldsymbol{X}_i} W \left( \boldsymbol{X}_{i}^{n+\frac{1}{2}}\right),
\end{equation}
where we take the gradient of the energy function $W$. Next, the structure force $\mathcal{F}_{i}^{n+\frac{1}{2}}$ will be spread out into the fluid through
\begin{equation}
\boldsymbol{g}^{n+\frac{1}{2}}(\boldsymbol{x})=\sum_{\boldsymbol{X} \in G_{s}} \mathcal{F}_{i}^{n+\frac{1}{2}} \delta_{\Delta}\left(\boldsymbol{x}-\boldsymbol{X}_{i}\right).
\label{eq:ibm_spread}
\end{equation}
For specific 2D simulations (e.g. sedimentation) an area scaling factor $A_b$ can be used in Eq. (\ref{eq:ibm_spread}) \cite{ghosh_stockie_2015}. Using the force density $\boldsymbol{g}^{n+\frac{1}{2}}$ in the LBM forcing scheme (e.g. Eq. (\ref{eq:guo_force})), we can then stream and collide using Eq. (\ref{eq:discrete_LBE}) and then calculate velocity $\boldsymbol{u}^{n+\frac{1}{2}}$ using Eq. (\ref{eq:macrosocpicvariables}). Finally, the solid position at time $n+1$ is updated as
\begin{equation}
\boldsymbol{X}_{i}^{n+1}=\boldsymbol{X}_{i}^{n}+ \sum_{\boldsymbol{x} \in g_{x}} \boldsymbol{u}^{n+\frac{1}{2}} \delta_{\Delta}\left(\boldsymbol{x}-\boldsymbol{X}^{n+\frac{1}{2}}_{i}\right) .
\label{eqn:final_position}
\end{equation}
We recall that we use $\Delta t=1$ and $\Delta x=1$ in the above equations.
\subsubsection{Lubrication corrections}
When the gap between particles or between particles and the wall is equal to or less than the resolution of the discretization, the fluid flow can no longer be resolved in this region. In most cases this happens when the gap is less than one lattice spacing. To solve this, there are two alternative solutions. The first one is to use a refined mesh for the entire fluid system (or just near the region in question if possible). This would lead to a more expensive simulation. Depending on the problem a slightly finer mesh might prove to be enough. The second approach is to introduce a lubrication force that mimics the repulsive force of the fluid being squeezed out of the gap. This second approach is computationally more manageable. This idea has been incorporated into the LBM \cite{ladd_lattice-boltzmann_2001,ladd_sedimentation_1997-2, nunes_dynamical_2023}, using lubrication theory. The lubrication force between two rigid spheres is
\begin{equation}
\boldsymbol{F}_{i j}^{\mathrm{lub}}=-\frac{3 \pi \mu r}{\mathrm{~s}} \hat{\boldsymbol{x}}_{i j} \hat{\boldsymbol{x}}_{i j} \cdot\left(\boldsymbol{u}_{i}-\boldsymbol{u}_{j}\right),
\label{eqn:lubrication_force}
\end{equation}
$\mu$ is the dynamic viscosity of the fluid, where $r$ is the radius of the spheres, $s$ is the dimensionless gap $s=R / r-2$ where $R$ is the central distance between two spheres. $\boldsymbol{x}_{i j}$ is the position vector difference between sphere $i$ and $j$, defined as $\boldsymbol{x}_{i j}=\boldsymbol{x}_{i}-\boldsymbol{x}_{j}$, $\hat{\boldsymbol{x}}_{i j}$ is the unit vector. $\boldsymbol{u}$ is the sphere's velocity. Equation (\ref{eqn:lubrication_force}) can also be extended to the case where a sphere approaches a stationary wall or object by setting $\boldsymbol{u}_{j}=0$.

For deformable bodies, a repulsive force between any two nodes of two meshes in close proximity is sufficient \cite{Gross_Krueger_Varnik_2014}. This force is zero for node-to-node distances larger than one lattice constant and increases as $1 / r^2$ at smaller distances
$$
\boldsymbol{g}_{i j}^{\mathrm{int}}=\left\{\begin{array}{rll}
-\kappa_{\mathrm{int}}\left(d_{i j}^{-2}-1\right) \frac{\boldsymbol{d}_{i j}}{d_{i j}} & \text { for } & d_{i j}<1, \\
0 & \text { for } & d_{i j} \geq 1
\end{array}\right.
$$
where $\boldsymbol{d}_{i j}$ is the distance between the two nodes $i$ and $j$, $\kappa_{\text {int }}$ is a constant regulating the strength and $\boldsymbol{g}_{j i}^{\mathrm{int}}=-\boldsymbol{g}_{i j}^{\text {int }}$.

\subsubsection{Other coupling schemes}

Here we discuss only the no-slip boundary conditions between fluid and structure, motivated by Ref. \cite{tan2015lattice}, which consists of imposing velocity and force continuity along the interface
\begin{equation}
\begin{gathered}
u_{f}=u_{s} \quad \text { on } \Upsilon \\
\sigma_{i j}^{f} n_{j}=\sigma_{i j}^{s} n_{j} \quad \text { on } \Upsilon
\end{gathered}
\label{eq:stressIntegration_continiuity}
\end{equation}
where $u_{f}$ and $u_{s}$ are the fluid and solid velocity, $\Upsilon$ is the fluid-solid boundary and $\sigma_{i j}$ is the stress tensor with superscript $f$ and $s$ meaning fluid and solid, respectively. $n_{j}$ is the surface normal. In the IBM, the solid velocity is interpolated from the fluid, as seen in Eq. (\ref{eqn:velocity_interpolation}), while the force is spread into the fluid, as shown in Eq. (\ref{eqn:force_distribution}). Alternatively, the inverse method can be used \cite{kwon_development_2006,kollmannsberger_fixed-grid_2009}. In this approach, we initially apply the fluid stress to the structure, subsequently solve for the structural response, and ultimately enforce the structural velocity as a boundary condition on the fluid. This technique is referred to as the stress integration approach \cite{kollmannsberger_fixed-grid_2009, macmeccan_simulating_2009}. The total force acting on the structure is
\begin{equation}
T_{i}=\int \sigma_{i j} n_{j} \mathrm{d}A
\end{equation}
where $\mathrm{d}A$ is the differential area over the interface. The stress tensor $\sigma_{i j}$ is given by
\begin{equation}
\sigma_{i j}=-p \delta_{i j}+\rho \nu\left(u_{i, j}+u_{j, i}\right) \ \ .
\label{eq:stress_tensor}
\end{equation}
The pressure term $p$ is usually assumed to follow an ideal gas law $p = \rho c_{s}^2$. Following Ref. \cite{kruger_2009}, the deviatoric shear stress in the LBM $\tau_{i j}:=\rho \nu\left(u_{i, j}+u_{j, i}\right)$ can be evaluated as
\begin{equation}
\tau_{i j}=-\left(1-\frac{\omega}{2}\right) \sum_{\alpha}\left(\xi_{\alpha i} \xi_{\alpha j}-\frac{\delta_{i j}}{D} \boldsymbol{\xi}_{\alpha} \cdot \boldsymbol{\xi}_{\alpha}\right) f_{\alpha}^{n e q} \Delta t
\label{eq:deviatoric_stress}
\end{equation}
where $\omega=\frac{\Delta t}{\tau}, \tau$ is the relaxation time for the LBM. $\alpha$ represents the discrete lattice speed, $f_{\alpha}^{n e q}$ is the non-equilibrium part of the density distribution defined as $f_{\alpha}^{n e q}=f_{\alpha}-f_{\alpha}^{eq}$ with $f_{\alpha}^{eq}$ calculated from Eq. (\ref{eqn:eq_distribution}). $\boldsymbol{\xi}_{\alpha}$ is the discretized velocity vector. Using the stress computed from the LBM through Eqs. (\ref{eq:stress_tensor}) and (\ref{eq:deviatoric_stress}), the force exerted on the structure can be calculated by taking the dot product between the surface normal and the stress tensor, using the second equation in (\ref{eq:stressIntegration_continiuity}). 

In soft matter systems, the flows are often at low Reynolds numbers. Therefore, it is appropriate to Stokes friction force to couple the immersed solids and fluids \cite{ahlrichs_lattice-boltzmann_1998,ahlrichs_simulation_1999,hsu_migration_2010,holm_lattice_2009}. In Ref. \cite{ahlrichs_simulation_1999}, polymer monomers are treated as points with a friction force
\begin{equation}
F_\zeta=-\zeta\left(u_{s}-u_{f}\right),
\end{equation}
where $\zeta$ is the friction coefficient, which might differ from the one provided by Einstein's diffusion relation. $u_{s}, u_{f}$ are solid point and local fluid velocities, respectively. Ref. \cite{ahlrichs_lattice-boltzmann_1998}notes that $u_{f}$ can be estimated through interpolation from the nearest neighbouring fluid nodes. The solid will experience a friction force $F_\zeta$, among other forces such as elastic, to determine its motion. Notice that in order to conserve momentum, a force of equal magnitude but opposite direction $-F_\zeta$ is applied to the neighbouring fluid nodes. For example, Ref. \cite{ahlrichs_lattice-boltzmann_1998,ahlrichs_simulation_1999} uses a force density $-F_\zeta / \delta x^{3}$ . Tuning might be necessary for an appropriate choice of the friction coefficient to reduce the effect of slip.

Lastly, while in this section we focus on the simulation of membrane-bound particles such as capsules using primarily the IBM, recent works have also used the IBM to simulate droplets \cite{li2019finite,pelusi2023sharp,taglienti2023reduced}.


\section{Fluid-Fluid based methods}
\label{sec_fluid_fluid}
This section gives an overview of widely multicomponent (different fluids) models for studying droplets and emulsions using the LBM as seen in Ref. \cite{succi_lattice_2001,kruger_lattice_2017,jackson2021droplet,shardt2014simulations}. These have applications in several fields, such as for example in injet printing \cite{jackson2019droplet}. The LBM facilitates the relatively seamless integration of inter-particle forces, forming interfaces between components. However, these forces are often simplified to save time in numerical simulations. Various models have been modified and improved to explore different parameters and reduce issues. Despite these efforts, no single model is perfect for all situations. First, we present the multicomponent formulation of each method and then we explain how frustrated coalescence can be achieved. 

\subsection{Colour Modelling}

Colour modelling was among the early multicomponent models developed as seen in Ref.~\cite{rothman1988a}. This model is an extension of a basic lattice-gas cellular automaton (LGCA) that handles two immiscible components with surface tension. Instead of the usual indistinguishable particles, it uses two distinct particles labelled as red and blue. The model follows rules similar to the standard LGCA, ensuring that collisions maintain the count of red and blue particles. This modification enables the simulation of immiscible fluids and incorporates surface tension effects within the LGCA framework. Yet, additional rules are introduced to promote the clustering of similar colours and generate surface tension effects. Following this, the model was integrated into the LBM in Ref. \cite{gunstensen1991a} and later, in Ref. \cite{grunau1993a} the model underwent changes to accommodate variations in viscosity and density ratios. Another key development was proposed in Ref.~\cite{latva-kokko2005a}, namely, to add a recolouring step which significantly reduces the computational requirements while minimising the spurious currents and removing the lattice pinning effect~\cite{Connington2012, PhysRevE.73.047701}. The model works by introducing the distribution function for each of the fluid components $f_{\alpha,k}(\boldsymbol{x}, t)$, where $k$ is the component index. The discrete LBM equation is given by
\begin{equation}
f_{\alpha,k}\left(\boldsymbol{x}+\boldsymbol{\xi}_{\alpha} \Delta t, t+\Delta t\right)=f_{\alpha,k}(\boldsymbol{x}, t)+\Omega_{\alpha,k}(\boldsymbol{x}, t),
\end{equation}
where $\Omega_{\alpha,k}$ is the multistage collision operator expressed as
\begin{equation}
\Omega_{\alpha,k}=\left[\Omega_{\alpha,k}^{(1)}+\Omega_{\alpha,k}^{(2)}\right] \Omega_{\alpha,k}^{(3)} \ \ .
\end{equation}
Each has a distinct aim:

\begin{itemize}

  \item $\Omega_{\alpha,k}^{(1)}$ is the normal single-phase collision operator for each component;

  \item $\Omega_{\alpha,k}^{(2)}$ is the perturbation operator, generating the interfacial tension;

  \item $\Omega_{\alpha,k}^{(3)}$ is the recolouring operator responsible for phase separation.

\end{itemize}
Macroscopic quantities are obtained by calculating the standard moments of $f_{\alpha,k}(x, t)$ :
\begin{equation}
\begin{aligned}
\rho_{k} &=\sum_{\alpha} f_{\alpha,k}, \\
\rho \boldsymbol{u} &=\sum_{\alpha, k}  \boldsymbol{\xi}_{\alpha} f_{\alpha,k},
\end{aligned}
\end{equation}
where $\rho=\Sigma_{k} \rho_{k}$. Consequently, the velocity $\boldsymbol{u}$ is referred to as the colour-blind velocity. The perturbation step, responsible for creating surface tension at the interface between two fluids i.e. components, is expressed as:
\begin{equation}
\Omega_{\alpha,k}^{(2)}=\frac{A_{k}}{2}|\nabla \rho|\left[w_{\alpha} \frac{\left(\boldsymbol{\xi}_{\alpha} \cdot \nabla \rho\right)^{2}}{|\nabla \rho|^{2}}-B_{\alpha}\right],
\end{equation}
where $\nabla \rho$ is the colour gradient, $A_{k}$ regulates the intensity of the surface tension and $B_{\alpha}$ are constants unique to the DNQn set. For the D2Q9 velocity set, $B_{\alpha}$ are given by
\begin{equation}
\begin{aligned}
&B_{0} =-\frac{\chi}{3 \chi+6} c^{2}, \quad B_{1, \ldots, 4}=\frac{\chi}{6 \chi+12} c^{2} \quad \text { and } \\
&B_{5, \ldots, 8} =\frac{1}{6 \chi+12} c^{2} .
\end{aligned}
\end{equation}
Likewise, for the D3Q19 velocity set, $B_{\alpha}$ are given by
\begin{equation}
\begin{aligned}
&B_{0} =-\frac{2+2 \chi}{3 \chi+12} c^{2}, \quad B_{1, \ldots, 6}=\frac{\chi}{6 \chi+24} c^{2} \text { and } \\
&B_{7, \ldots, 18} =\frac{1}{6 \chi+24} c^{2},
\end{aligned}
\end{equation}
where $\chi$ is a free parameter: Values for $B_{\alpha}$ and $\chi$ are derived in Ref. \cite{Leclaire_Reggio_Trepanier_2012}. Calculation of partial derivatives is necessary for $\nabla \rho$ and it has been shown \cite{liu2012a} that the following isotropic
central difference contributes to enhancing numerical stability and lowering the discretization error,
\begin{equation}
\partial_{j} \rho(\boldsymbol{x}, t)  \approx \frac{1}{c_{s}^{2} \Delta t} \sum_{\alpha} w_{\alpha} \rho\left(\boldsymbol{x}+\boldsymbol{\xi}_{\alpha} \Delta t, t\right) \xi_{\alpha j}, 
\end{equation}
where $j$ runs through the spatial coordinates. Lastly, one just needs to do the  recolouring. In Ref. \cite{rothman1988a} this involves the minimisation of the work, $W$, which is given by
\begin{equation}
W\left(f^{r}, f^{b}\right)=-\nabla \rho \cdot \boldsymbol{q}\left(f^{r}, f^{b}\right),
\end{equation}
where $\boldsymbol{q}$ is the local or colour flux
\begin{equation}
\boldsymbol{q} \left(f^{r}, f^{b}\right)=\sum_{\alpha} \boldsymbol{\xi}_{\alpha}\left(f_{\alpha}^{r}-f_{\alpha}^{b}\right).
\end{equation}
Nevertheless, this recolouring method may lead to pinning. This was corrected in Ref. \cite{latva-kokko2005a} by enabling the blending of red and blue components through symmetric colour distribution concerning the colour gradient. The particle distributions after recolouring are given as
\begin{equation}
\begin{aligned}
{\Omega_{\alpha}^r}^{(3)} (f_{\alpha}^{r}) &=\frac{\rho_{r}}{\rho} f_{\alpha}^{\prime}+\mathcal{B} \frac{\rho_{r} \rho_{b}}{\rho^{2}} f_{\alpha}^{e q}(\rho, \boldsymbol{u}=0) \cos \left(\theta_{\alpha}\right), \\
{\Omega_{\alpha}^b}^{(3)} (f_{\alpha}^{b}) &=\frac{\rho_{b}}{\rho} f_{\alpha}^{\prime}-\mathcal{B} \frac{\rho_{r} \rho_{b}}{\rho^{2}} f_{\alpha}^{e q}(\rho, \boldsymbol{u}=0) \cos \left(\theta_{\alpha}\right),
\end{aligned}
\label{eq:recolouring_stepRK}
\end{equation}
where $\mathcal{B} \in(0,1) $ is used to manage the thickness of the interface, $f_{\alpha}^{\prime}=\Sigma_{k} f_{\alpha, k}^{\prime}$ where $f_{\alpha, k}^{\prime}$ is the post-collision state after $\Omega_{\alpha,k}^{(1)}$ and $\Omega_{\alpha,k}^{(2)}$ have been applied. 
The term $\cos \left(\theta_{\alpha}\right)$ is conveniently expressed as
\begin{equation}
\cos \left(\theta_{\alpha}\right)=\frac{\boldsymbol{\xi}_{\alpha} \cdot \nabla \rho}{|\boldsymbol{\xi}_{\alpha}||\nabla \rho|} .
\end{equation}
The total zero-velocity equilibrium distribution function
$f_{\alpha}^{e q}(\rho, \boldsymbol{u}=0)=\sum_k f_{\alpha,k}(\rho, \boldsymbol{u}=0)^{e q}$ is given by
\begin{equation}
f_{\alpha}^{eq}=\rho\left(\phi_{\alpha}+w_{\alpha}\left[3 \boldsymbol{\xi}_{\alpha} \cdot \boldsymbol{u}+\frac{9}{2}\left(\boldsymbol{\xi}_{\alpha} \cdot \boldsymbol{u}\right)^2-\frac{3}{2}(\boldsymbol{u})^2\right]\right)
\end{equation}
where the weights $\phi_{\alpha}$ (along with $w_{\alpha}$) are lattice dependent \cite{Reis_Phillips_2007,Leclaire_Parmigiani_Malaspinas_Chopard_Latt_2017}. Further modification of the recolouring step in Eq. (\ref{eq:recolouring_stepRK}) to improve spurious velocities, lattice pinning and convergence was done first in Ref. \cite{Reis_Phillips_2007}. Furthermore, $\theta_{\alpha}$ is the angle between $\nabla \rho$ and $\boldsymbol{\xi}_{\alpha}$ . 
Finally, the streaming step is carried for each colour.

The advantages of the colour model include its ease of extension beyond two components. Moreover, it provides free control over both the surface tension and thus, the interface thickness. Recent advancements have enabled the simulations of fluids with high density \cite{ba2016multiple} and viscosity ratios \cite{liu2016lattice}. One possible limitation, contingent on the modelling needs, is the absence of a direct connection to thermodynamics. Another downside involves the creation of undesired velocities at curved interfaces, a phenomenon commonly observed in various other fluid-fluid models.

\subsubsection*{Frustrated coalescence}
In colour modelling, a repulsive force has been used to prevent the coalescence of immiscible droplets \cite{Montessori_Lauricella_Succi_2019,Montessori_Lauricella_Tirelli_Succi_20192,Montessori_Tiribocchi_Bonaccorso_Lauricella_Succi_2020}. This repulsive force term (operating exclusively on the fluids interfaces) can be introduced to account for any near-contact forces acting between droplets that are closer than the lattice spacing
\begin{equation}
\boldsymbol{F}_{\text {rep }}=-A_{ h}[ h(\boldsymbol{x})] \boldsymbol{n} \delta_I,
\end{equation}
where $A_{ h}[ h(\boldsymbol{x})]$ is the parameter controlling the strength of this repulsive force, $\boldsymbol{n}$ is a unit normal to the interface and $\delta_I=\frac{1}{2}|\nabla \phi|$ is a function proportional to the phase field $\phi=\frac{\rho^1-\rho^2}{\rho^1+\rho^2}$ which confines the repulsive force to the interface. A graphical representation of this force is shown in Fig.~\ref{fig:colourmodelling_repulsion}.  $A_{ h}[ h(\boldsymbol{x})]$ can be set equal to a constant $A$ if $ h \leq  h_{\min }$, it decays as $ f^{-3}$ if $ h_{\min }< h \leq h_{\max }$ and it is equal to zero if $ h>  h_{\max }$. For example in Ref. \cite{Montessori_Tiribocchi_Bonaccorso_Lauricella_Succi_2020}, $ h_{\min }=2$ and $ h_{\max }=4$ lattice spacing. Despite the possibility of other functional forms, this one is generally adequate to prevent droplet coalescence and, more significantly, to accurately describe the physics at various length scales. The repulsive force leads to
\begin{align}
\frac{\partial (\rho \boldsymbol{u})}{\partial t}+\nabla \cdot \left( \rho \boldsymbol{u} \boldsymbol{u} \right)&=-\nabla p+\nabla \cdot\left[\rho v\left(\nabla \boldsymbol{u}+\nabla \boldsymbol{u}^{\mathrm{T}}\right)\right] \nonumber\\&+\nabla \cdot(\boldsymbol{\Sigma}+\pi \boldsymbol{l}),
\end{align}
which is the NS momentum equation for a multicomponent system, with a
surface-localised repulsive term, expressed through the potential function $\nabla \pi$. Including extra forces to frustrate coalescence can in principle be used for any multicomponent method, as will be discussed in the following sections.

\begin{figure}
\center
\includegraphics[width=0.38\textwidth]{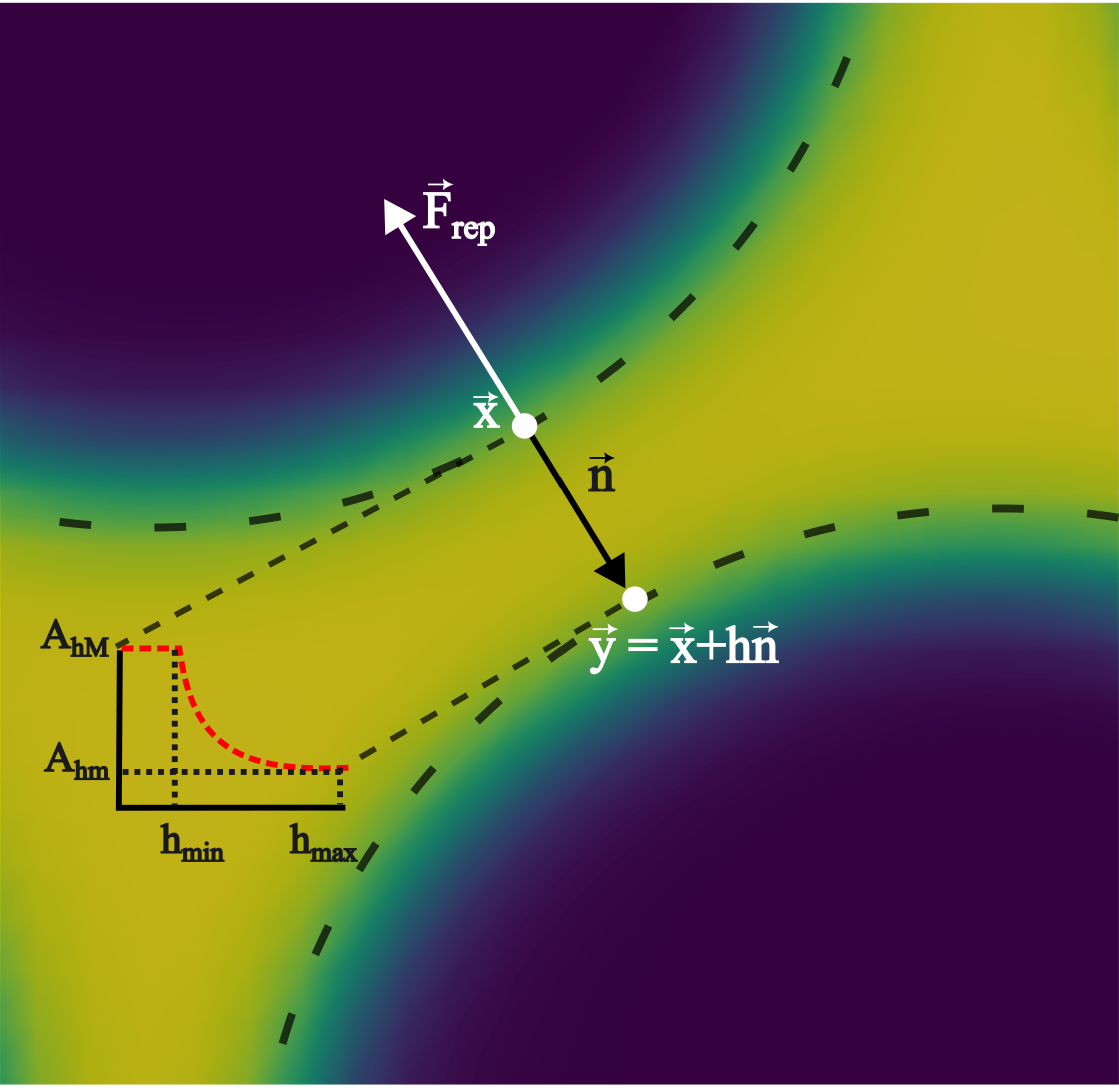}
\caption{ Near contact forces representation. Mesoscale modelling of the interactions at the point of contact of two immiscible fluid droplets. The repulsive force is denoted by $\boldsymbol{F}_{\text {rep }}$, and the unit vector $\boldsymbol{n}=-\nabla \rho /|\nabla \rho|$ is perpendicular to the fluid interface. The dotted line follows the fluid interface outermost boundary, and the vectors $\boldsymbol{x}$ and  $\boldsymbol{y}$ represent the coordinates taken there. The repulsive parameter maximum and minimum values are $A_{ h M}$ and $A_{ h m}$, respectively. The colours represent the density field i.e. yellow is the continuum fluid and blue is the dispersed (droplets) fluid.}
\label{fig:colourmodelling_repulsion}
\end{figure}

\subsection{Free-Energy Modelling}
\label{sec:freenergy}
We introduce a straightforward free-energy LBM for binary liquid mixtures \cite{swift1996a,Briant_Yeomans_2004}. In this method, discrete density distributions $f_{k,\alpha}$ for each $k$-component model the hydrodynamics and the evolution of a phase field, respectively. This approach utilises a Cahn-Hilliard phase field method to represent a scalar determining the fluid's composition. We will describe a model where both the fluid flow and evolution of the scalar phase field are solved with the LBM. However, it is also possible to use a hybrid scheme such that the fluid flow is solved with LBM while the phase field is solved with finite differences which leads to reduced memory requirement and usually a marginal loss of numerical stability~\cite{PhysRevE.76.031921, Coelho_Araujo_Telo_Gama_2021, C4SM01382D, C4SM00894D}. The scalar not only moves with the flow but also undergoes diffusion to minimise a free-energy functional. The functional includes two components: a double-well potential leading to phase separation and an energy penalty discouraging composition gradients, giving tension to interfaces. The thickness of the interface
and its interfacial tension can be specified independently (in terms of $\alpha$ and $\kappa_\phi$ as discussed below Eq.~\eqref{eq:Marenduzzo_freeenergydensity}) which is an important
advantage over the pseudopotential models described in the next section.

The distribution function $f_{k,\alpha}$ evolves according to

\begin{multline}
f_{k,\alpha}\left(\boldsymbol{x}_{\alpha}+\boldsymbol{\xi}_{\alpha} \Delta t, t+\Delta t\right)=\\f_{k,\alpha}\left(\boldsymbol{x}_{\alpha}, t\right)-\frac{\Delta t}{\tau_k}\left[f_{k,\alpha}\left(\boldsymbol{x}_{\alpha}, t\right)-f_{k,\alpha}^{eq}\left(\boldsymbol{x}_{\alpha}, t\right)\right]+\mathcal{F}_{k,\alpha}\Delta t,
\label{eq:}
\end{multline}
where $\tau_k$ specify the relaxation rates. By relating macroscopic values to the density distributions and appropriately choosing the equilibrium distributions $f_{k,\alpha}^{e q}$, the simulations model the required continuum equations. For the $f_{k,\alpha}$ field, the macroscopic density and momentum are
\begin{equation}
\rho_{k}=\sum_{\alpha=0} f_{k, \alpha}, \quad \rho_{k} \boldsymbol{u}_{k}=\sum_{\alpha=0} f_{k, \alpha} \boldsymbol{\xi}_{\alpha}.
\end{equation}
and the equilibrium distribution is
\begin{equation}
f_{k,\alpha}^{e q}=A_{\alpha}+B_{\alpha} \boldsymbol{u} \cdot \boldsymbol{\xi}_{\alpha}+C_{\alpha} \boldsymbol{u} \cdot \boldsymbol{u}+D_{\alpha}\left(\boldsymbol{u} \cdot \boldsymbol{\xi}_{\alpha}\right)^2+G_{\alpha, x y} \xi_{\alpha,x} \xi_{\alpha, y},
\end{equation}
where index notation has been used for the last term and summation over repeated $xy$ indices is implied. The coefficients $A_{\alpha}, B_{\alpha}, C_{\alpha}, D_{\alpha}$ and $G_{\alpha}$ needs to adhere to conservation constraints, yet these constraints do not uniquely define the coefficients. It is possible to use coefficients that reduce spurious currents \cite{Pooley_Furtado_2008}. For a D2Q9 velocity set, the full list of coefficients is given in Ref. \cite{huang_multiphase_2015} and details for construction on a D3Q19 velocity set are given in Ref. \cite{Pooley_Furtado_2008}.

For the phase field, the scalar $\phi$ specifies the composition of the fluid, and it varies between -1 (continuum phase) and 1 (droplet phase). It is determined from the density distribution by
\begin{equation}
    \phi=\sum_{k,\alpha} f_{k,\alpha} \ \ .
\end{equation}

The continuum equation for $\phi$ is the Cahn-Hilliard one
\begin{equation}
\frac{\partial \phi}{\partial t}+\nabla \cdot(\boldsymbol{u} \phi)=M \nabla^2 \mu,
\end{equation}
where $M$ is the mobility of the chemical potential $\mu$. This diffusivity is determined by the relaxation time $\tau_\phi$ and a free parameter $P$ according to $M=P\left(\tau_g-\frac{1}{2}\right)$, while the chemical potential is determined by the free-energy of the system. The free-energy functional $\mathfrak{F}[\phi(\boldsymbol{}{x})]$ is \cite{Briant_Yeomans_2004}
\begin{equation}
\mathfrak{F}=\int_V\left[\frac{1}{c_{s}^{2}} \rho \ln \rho+\frac{1}{2} \phi^2\left(-A+\frac{B}{2} \phi^2\right)+\frac{\kappa_{\phi}}{2}(\nabla \phi)^2)\right] \mathrm{d} V
\end{equation}
The first term gives an ideal gas equation of state, the second term is a double-well potential that causes phase separation with minima at compositions $\phi_0= \pm \sqrt{\frac{A}{B}}$, and the third term sets the interfacial tension by associating an energy penalty for changes in $\phi$ with an elastic constant $\kappa_{\phi}$. The parameters $A$ and $B$ specify the shape of the double-well potential. Two symmetrical phases with $\phi_0= \pm 1$, are obtained when $A=B$. The strength of the free energy resulting from concentration gradients is governed by the parameter $\kappa_{\phi}$. The chemical potential is then given by
\begin{equation}
\mu_{\phi}=\frac{\delta \mathfrak{F}}{\delta \phi}=-A \phi+B \phi^3-\kappa_{\phi} \nabla^2 \phi
\end{equation}
The one-dimensional steady-state solution for $\phi$ between two infinite domains offers crucial insights into the interface, particularly regarding its thickness and excess free energy. The solution is as follows \cite{Briant_Yeomans_2004}
\begin{equation}
\phi(x)=\phi_0 \tanh \frac{x}{\ell_{\phi}},
\end{equation}
where $\ell_\phi=\sqrt{2 \kappa_\phi / B}$ is the interfacial thickness of the droplets. The free energy model is also used to study active droplets \cite{Tiribocchi_Durve_Lauricella_Montessori_Marenduzzo_Succi_2023,Coelho_Araujo_Telo_Gama_2021,PhysRevResearch.5.033165}.

\subsubsection*{Frustrated coalescence}

Literature on frustrated coalescence using the free energy LBM is scarce. We present a method used by Ref. \cite{Foglino_Morozov_Henrich_Marenduzzo_2017} to study the shear thinning of immiscible droplets in a 2D channel driven by flow. It tracks two fields: the phase-field variables representing each droplet's density $\phi_i$ where $i=1, \ldots, N$ is the number of droplets and the fluid velocity field $\boldsymbol{}{u}$. The free energy is given by:
\begin{equation}
\mathfrak{F}=\int_V\left[ \frac{\alpha}{4} \sum_i^N \phi_i^2\left(\phi_i-\phi_0\right)^2+\frac{\kappa_{\phi}}{2} \sum_i^N\left(\nabla \phi_i\right)^2+\epsilon \sum_{i, j, i<j} \phi_i \phi_j \right] \mathrm{d} V ,
\label{eq:Marenduzzo_freeenergydensity}
\end{equation}
where the first term represents the double well potential which ensures droplet stability. It yields two minima for $\phi_i=\phi_0$ and $\phi_i=0$, which represent the inside and outside region of the $\mathrm{i}$-th droplet, respectively. The droplet deformability properties are determined by their surface tension $\Gamma=$ $\sqrt{8 \kappa_{\phi} \alpha / 9}$, and can therefore be tuned by changing the value of $\kappa_{\phi}$ in Eq.~(\ref{eq:Marenduzzo_freeenergydensity}). The parameters $\kappa_{\phi}$ and $\alpha$ also determine the interfacial thickness of the droplets  $\ell_{\phi}=\sqrt{2 \kappa_{\phi} / \alpha}$. The third, final term describes a soft repulsion pushing the droplets apart when they overlap, therefore preventing coalescence. The strength of this repulsion is regulated by the value of the positive constant $\epsilon$. The dynamics of $\phi_i$ is given by
\begin{equation}
\frac{\partial \phi_i}{\partial t}+\nabla \cdot\left(\boldsymbol{u} \phi_i\right)=M \nabla^2 \mu_i,
\label{eq:Cahn-Hilliard}
\end{equation}
where $M$ is the mobility and $\mu_i=\partial \mathfrak{F} / \partial \phi_i-\partial_\alpha \partial \mathfrak{F} / \partial\left(\partial_\alpha \phi_i\right)$ is the chemical potential of the $i$-th droplet. Equations (\ref{eq:Marenduzzo_freeenergydensity}) and (\ref{eq:Cahn-Hilliard}) are solved using a combination of the LBM (for the fluid flow) and finite differences (for the phase field) similar to Ref. \cite{swift1996a,Tiribocchi_Stella_Gonnella_Lamura_2009}.
This may be used for a small number of droplets. As the number of times that Eqs. (\ref{eq:Marenduzzo_freeenergydensity}) and (\ref{eq:Cahn-Hilliard}) are solved increases the computational cost rapidly as the number of droplets increases. This method is also useful to simulate droplets inside droplets \cite{Tiribocchi_Montessori_Lauricella_Bonaccorso_Succi_Aime_Milani_Weitz_2021}.

\begin{figure}
\center
\includegraphics[width=0.35\textwidth]{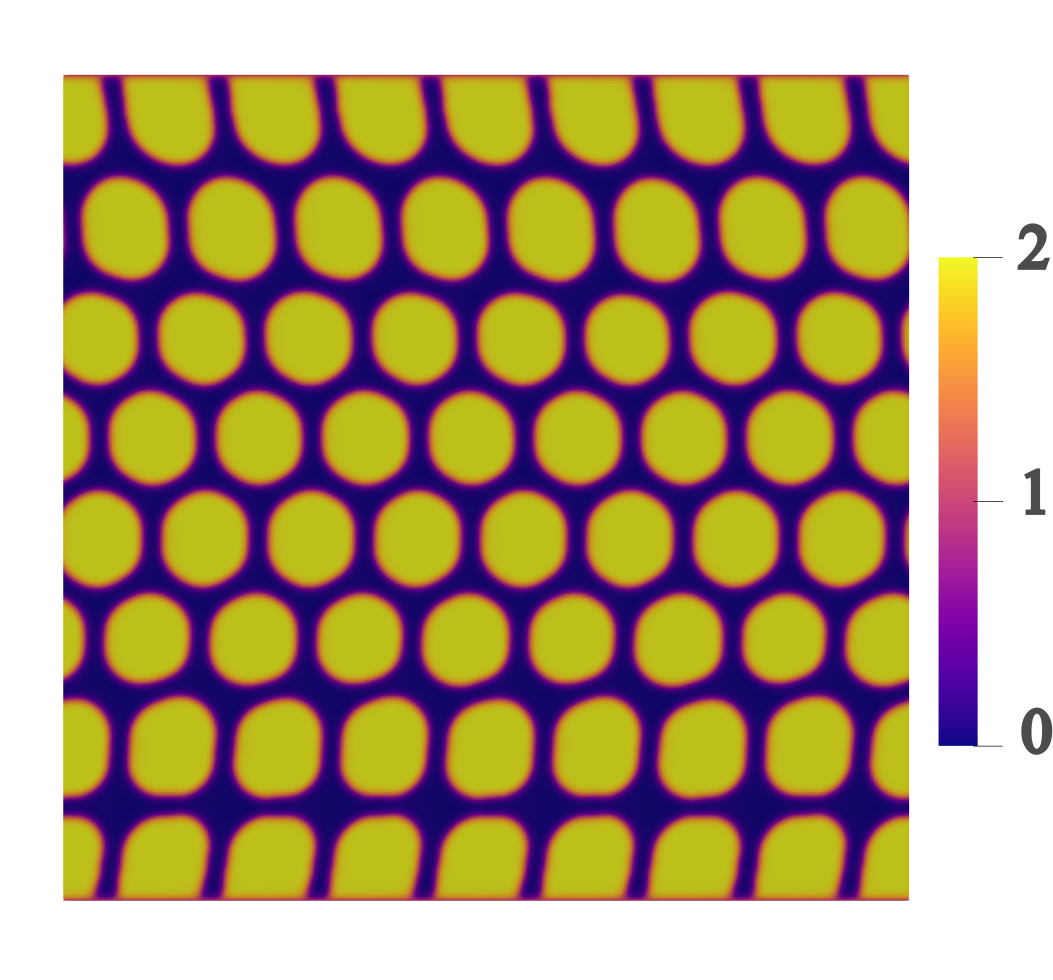}
\caption{Droplets moving in a channel. This was obtained using a free energy LBM and solving Eq.~(\ref{eq:Cahn-Hilliard}). Colour depicts $\sum_i \phi_i$ which is $\approx 2$ for droplets (yellow) and $\approx 0$ for the background fluid (black).}
\end{figure}

\subsection{Pseudopotential Method}
\label{sec:pseudopotential}
The pseudopotential (or Shan-Chen models) first introduced in Ref. \cite{shan1993a, shan1994a} as a way to simulate multicomponent and multiphase flows. It works by adding inter-particle interactions resulting in the separation of components. Some instances of applications of pseudopotential models include gravity driven droplets in confined environments \cite{kang2005a} and transport of H2O in air \cite{bao2013a,stiles2016a}. Different fluids in a multicomponent mixture have different molecular forces. The force acts on pairs of molecules located at $\boldsymbol{x}$ and $\tilde{\boldsymbol{x}} \neq \boldsymbol{x}$. It is also assumed that a higher density $\rho(\boldsymbol{x})$ of molecules leads to larger forces. The total force is given as an integral over all possible effective interactions as
\begin{equation}
\boldsymbol{F}(\boldsymbol{x})=-\int(\tilde{\boldsymbol{x}}-\boldsymbol{x}) G(\boldsymbol{x}, \tilde{\boldsymbol{x}}) \psi(\boldsymbol{x}) \psi(\tilde{\boldsymbol{x}}) \mathrm{d}^{3} \tilde{\boldsymbol{x}},
\end{equation}
where $G(\boldsymbol{x}, \tilde{\boldsymbol{x}})$ strength of force between two fluid elements at $\mathbf{x}$ and $\mathbf{x}^\prime$ and $\psi(\boldsymbol{x})$ is the pseudopotential, which is a function of the fluid density. Assuming nearest-neighbour interactions at the mesoscopic scale, the effective force on the $k$th component is given by
\begin{equation}
\boldsymbol{F}_{k}^{r}(\boldsymbol{x})=-\psi_{k}(\boldsymbol{x}) \sum_{\tilde{k}} G_{\tilde{k} k} \sum_{i} w_{i} \psi_{\tilde{k}}\left(\boldsymbol{x}+\boldsymbol{\xi}_{i} \Delta t\right) \boldsymbol{\xi}_{i}
\label{eq:force_repulsive}
\end{equation}
where $G_{\tilde{k} k}$ is a $k \times k$ matrix for the molecular interaction strength between fluid components and its sign regulates whether the forces are attractive (negative) or repulsive (positive). Subscript $r$ stands for repulsive. To create immiscible droplets $\boldsymbol{F}_{r}^{k}$ needs to be repulsive. $k$ and $\tilde{k}$ are just two different components of the fluid system. For a multicomponent mixture (no multiphase), the diagonal components of the matrix are zero and $G_{AB}=G_{BA}$ for two components $A$ and $B$. $\psi_{k}$ is the pseudopotential (or effective density). The pseudopotential can take different forms. A commonly used form is
\begin{equation}
\psi_{k}(\boldsymbol{x},t)=\rho_{0}\left[1-\exp \left(\frac{\rho_k}{\rho_{0}}\right)\right],
\label{eq:pseudopotential}
\end{equation}
where $\rho_{0}$ is a freely chosen reference density, usually unity. $\psi_{k}(\boldsymbol{x},t) = \rho_{k} $ can also be used and it is an approximation of Eq.~(\ref{eq:pseudopotential}) at low-density. However, at higher densities, this assumption will lead to numerical instabilities~\cite{sbragaglia2007a}. In the absence of attraction between components, the pseudopotential can often be reduced to the physical density for simplicity. Nevertheless, using Eq.~(\ref{eq:pseudopotential}) instead of $\psi_{k}(\boldsymbol{x},t) = \rho_{k} $ leads to improved numerical stability. 

The pseudopotential model allows for multicomponent multiphase (MCMP) simulations, but, for simplicity, we do not present the variants of pseudopotential MCMP LBM. In what follows, we focus only on the multicomponent variant as it is the most widely used method for immiscible droplets. In Ref. \cite{shan1993a, shan1994a}, they proposed a simple forcing scheme to include these interactions. Similar to the body force, it is also possible to incorporate the effective force of the interaction potential
in the fluid velocity field. Nevertheless, for a multicomponent system, we introduce $\boldsymbol{u}^{eq}$ which is a combined velocity (different from the physical velocity) of different components given by:
 \begin{equation}
\boldsymbol{u}^{eq}=\frac{\sum_{k} \frac{\rho_{k} \boldsymbol{u}_{k}}{\tau_{k}}}{\sum_{k} \frac{\rho_{k}}{\tau_{k}}} ,
\end{equation}
where $\rho_{k} \boldsymbol{u}_{k}$ is the $k$th component of momentum and $\tau_k$ is the relaxation time of the BGK collision operator for the component $k$. The barycentric velocity $\boldsymbol{u}$ of the fluid mixture, i.e., the physical velocity and the total density are given by
\begin{equation}
\boldsymbol{u}=\frac{\sum_{k} \rho_{k} \boldsymbol{u}_{k}}{\rho}, \quad \rho=\sum_{k} \rho_{k}. 
\end{equation}
Also, the density and momentum of each component can be calculated using,
\begin{equation}
\rho_{k}=\sum_{\alpha=0} f_{k, \alpha}, \quad \rho_{k} \boldsymbol{u}_{k}=\sum_{i=0} f_{k, \alpha} \boldsymbol{\xi}_{\alpha} + \frac{ \boldsymbol{F}_{k} }{2}.
\end{equation}

For a multi-component system, the time evolution of $f_{k, \alpha}\left(\boldsymbol{x},t\right)$ is given by the discrete Boltzmann equation for each component $k$ (similar to Eq. (\ref{eq:discrete_LBE})),
\begin{multline}
f_{k,\alpha}\left(\boldsymbol{x}_{i}+\boldsymbol{\xi}_{\alpha} \Delta t, t+\Delta t\right)=\\f_{k,\alpha}\left(\boldsymbol{x}_{i}, t\right)-\frac{\Delta t}{\tau_k}\left[f_{k,\alpha}\left(\boldsymbol{x}_{i}, t\right)-f_{k,\alpha}^{eq}\left(\boldsymbol{x}_{i}, t\right)\right]+\mathcal{F}_{k,\alpha}\Delta t .
\label{eq:discrete_LBE_multicomponent}
\end{multline}
The equilibrium distribution is given by
\begin{equation}
f_{k, \alpha}^{eq}=\rho_{k} w_{\alpha}\left[1+\frac{\boldsymbol{u}^{eq} \cdot \boldsymbol{\xi}_{\alpha}}{c_{s}^{2}}+\frac{\left(\boldsymbol{u}^{eq} \cdot \boldsymbol{\xi }_{\alpha}\right)^{2}}{2 c_{s}^{4}}-\frac{\left(\boldsymbol{u}^{eq}\right)^{2}}{2 c_{s}^{2}}\right] .
\label{eq:equillibirum_dist}
\end{equation}
The Guo forcing scheme~\cite{guo2002a} can be used to implement these forces acting in a fluid as it yields a viscosity-independent surface
tension \cite{Fei_Scagliarini_Montessori_Lauricella_Succi_Luo_2018}, particularly with the frustrated coalescence mechanism described in the next section. Thus similar to Eq. (\ref{eq:guo_force}),
\begin{equation}
\mathcal{F}_{k, \alpha}=\left(1-\frac{1}{2 \tau_{k}}\right) w_{\alpha}\left[\frac{\boldsymbol{\xi }_{\alpha}-\boldsymbol{u}^{e q}}{c_{s}^{2}}+\frac{\left(\boldsymbol{\xi }_{\alpha} \cdot \boldsymbol{u}^{e q}\right) \boldsymbol{\xi}_{\alpha}}{c_{s}^{4}}\right] \cdot \boldsymbol{F}_{k},
\label{eq:guo_force_component}
\end{equation}
where $\boldsymbol{F}_{k}$ is the total force acting on component $k$. In this case  $\boldsymbol{F}_{k} = \boldsymbol{F}_{k}^{r}$ . Different forcing schemes can be used. Additionally, as shown in the next section, the frustrated coalescence mechanism can be introduced as an extra force term. Although we described a model based on the BGK operator, extensions for MRT are possible~\cite{Chai2012}.

The pseudopotential model provides isotropy of the surface tension and spontaneous component
separation which improves the numerical efficiency of the simulation. The method is
successful because of its simplicity and flexibility in large-scale simulations of complex fluids. The implementation is relatively simple when compared to the other models reported so far. However, it also comes with some nonphysical features such as a diffusing interface and spurious currents~\cite{kruger_lattice_2017}. Some attempts have also been made to derive a thermodynamically consistent pseudopotential model \cite{li2014thermodynamic,huang2019thermodynamic,peng2021thermodynamically}. Moreover, it is possible to obtain an effective free energy in the pseudopotential models~\cite{Sbragaglia_2009, PhysRevE.84.036703}

\subsubsection*{Frustrated coalescence}
\label{subsec:pseudofrustrated}
To prevent coalescence a multi-range repulsive force is used which acts between the nearest neighbours and the next nearest neighbours \cite{Benzi_Sbragaglia_Scagliarini_Perlekar_Bernaschi_Succi_Toschi_2015,Fei_Scagliarini_Montessori_Lauricella_Succi_Luo_2018}. This force is used in addition to the repulsive component separation force $\boldsymbol{F}_{k}^{r}$. This means that in Eq. (\ref{eq:guo_force_component}), $\boldsymbol{F}_{k} = \boldsymbol{F}_{k}^{r} + \boldsymbol{F}_{k}^{c}$ where $\boldsymbol{F}_{k}^{c}$ is a competing force given by
\begin{equation}
\begin{split}
    \boldsymbol { F } _ { k } ^ { c } = - G _ { k , 1 } \psi _ { k } ( \boldsymbol { x } ) \sum _ { i = 0 } ^ { L1 } m _ { i } \psi _ { k } \left( \boldsymbol { x } + \boldsymbol { \xi } _ { i } \right) \boldsymbol { \xi } _ { i }\\ - G _ { k , 2 } \psi _ { k } ( \boldsymbol { x } ) \sum _ { j = 0 } ^ { L2 } n _ { j } \psi _ { k } \left( \boldsymbol { x } + \boldsymbol { \zeta } _ { j } \right) \boldsymbol { \zeta} _ { j },
    \end{split}
    \label{eq:competing_force}
\end{equation}
where $G _ { k , 1 }$ and $G _ {k , 2 }$ regulate the strength of force. It is a competing mechanism (superscript $c$) where the first term is an attractive force and the second term is a repulsive force between same components. This multi-range potential uses 2 different lattices $L1$ and $L2$. In 2D these are the D2Q9 and D2Q25 \cite{Benzi_Sbragaglia_Scagliarini_Perlekar_Bernaschi_Succi_Toschi_2015}, respectively while in 3D they are D3Q41 for $L1$ and D3Q39 for $L2$ \cite{chika}. In 2D, it is possible to completely separate the range of the forces i.e. the $L1$ lattice runs over only on the first belt of neighbours while $L2$ runs over on the second belt of neighbours. However, in 3D both lattices (D3Q41 and D3Q39) run over to the third belt of neighbours but on different lattice points which is sufficient to prevent droplet coalescence \cite{Silva_Coelho_daGama_Araujo_2023}. In the competing force $\boldsymbol{F}_{k}^{c}$, the attractive force must overcome the repulsive force to stabilise droplets, so we set $\left|G_{k, 1}\right|>\left|G_{k, 2}\right|$. To obtain the desired repulsive and attractive forces: $G_{k, 1}<0$, $G_{k, 2}>0$, and $G_{k, \overline{k}}>0$. For simplicity, we assume: $G_{A, 1}=G_{B, 1}$, $G_{A, 2}=G_{B, 2}$, $G_{A, B}=G_{B, A}$ for the two components $A$ and $B$. 

The computational cost does not increase with the number or size of the droplets by contrast to the colour gradient. Compared to the free energy model where each droplet is a different component, the pseudopotential model performs computationally better. This provides an efficient algorithm for simulating a large number of droplets \cite{coelho2023}. This is because the repulsion between droplets is based on the density of each LB lattice node. The extra cost comes from using a second lattice to calculate the force. However, this model suffers from spurious velocities caused by an imbalance between the forces. Using a higher order lattice for the streaming can help to reduce these velocities \cite{Silva_Coelho_daGama_Araujo_2023}. These velocities increase as the viscosity ratio deviates from unity or the surface tension increase. Proper and careful calibration of the $G$ parameters is necessary otherwise the droplet may increase or decrease in size. With this method, it is not possible to vary the surface tension and the interface thickness independently.

\begin{figure}
\center
\includegraphics[width=0.45\textwidth]{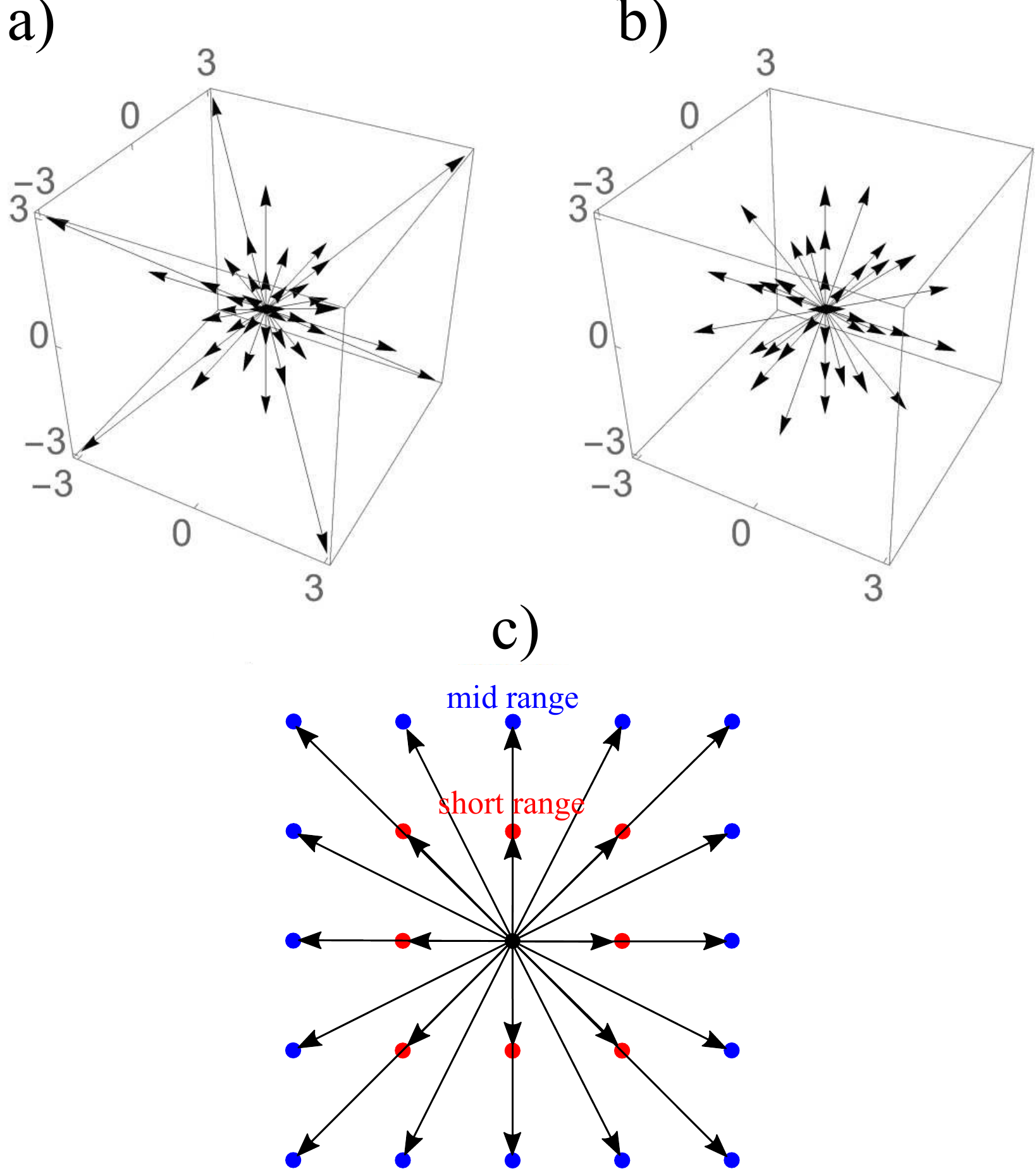}
\caption{(a) The D3Q41 lattice and (b) the D3Q39 lattice. The D3Q41 is a zero-one-three lattice meaning that it is analogous to the Brillouin zones (also referred to as lattice-belts, or lattice-stencils) covered by $\xi_\alpha$. The D3Q39 is a zero-one-two-three lattice. (c) The D2Q9 lattice for short-range nodes (red). The D2Q25 lattice for short and mid-range nodes (blue).}
\end{figure}

\section{Benchmarks}
\label{sec:benchmarks}

We now summarise a few simple benchmarks that can be used to test the model implementation in hydrostatic and hydrodynamic conditions. The benchmarks are for one capsule/droplet in shear flow, capsule sedimentation and a test of the Laplace law. We refer the reader to Ref. \cite{Fei_Scagliarini_Montessori_Lauricella_Succi_Luo_2018} for benchmarks on multiple droplets. 

\subsection{Capsule and droplet in shear flow}
Shear flow is a common validation flow to test for both droplets \cite{yiotis2007a,Soligo_Roccon_Soldati_2020} and capsules \cite{Eggleton_Popel_1998,macmeccan_simulating_2009} (fluid-structure interactions). Here we use capsules as a generic term for membrane-bound particles. This benchmark consists of placing a spherical (circular in 2D) droplet/capsule in shear flow (between two parallel walls with opposite velocities) and allowing it to deform until a steady state regime is reached \cite{Soligo_Roccon_Soldati_2020} (see Fig. \ref{fig:capsuelanddropletinshear}). In what follows, we have separated into two subsections the shear flow of a droplet and a capsule for clarity.

\begin{figure}
\center
\includegraphics[width=0.35\textwidth]{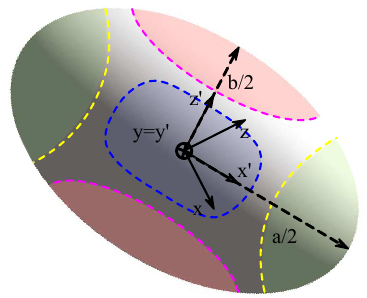}
\caption{The final steady-state shape of droplet/capsule with the major axis, $a$, the minor axis of deformation, $b$, and the third axis, $c$. An additional reference frame $\left(x^{\prime}, y^{\prime}, z^{\prime}\right)$ is defined, with the axes corresponding to the deformed droplet principal axes. The different regions of the droplet have been also highlighted for ease of reference: tips (green), belly (red) and sides (blue).}
\label{fig:capsuelanddropletinshear}
\end{figure}

\subsubsection{Droplet}
As a droplet deforms in shear flow, the final shape is a result of viscous forces competing with surface tension. The ratio of these forces is the Capillary number, 
\begin{equation}
    \mathrm{Ca} = \frac{u_w \mu}{\Gamma}\frac{R}{H},
\end{equation}
where $u_w$ is the wall velocity, $H$ is the distance between walls, $R$ is the radius of the droplet, $\mu$ the dynamic viscosity and $\Gamma$ is the surface tension. $\mathrm{Ca}$ is the main parameter that controls the deformation and final shape of the droplet. The final shape is evaluated using the Taylor deformation parameter given by
\begin{equation}
    D = \frac{a-b}{a+b}
\end{equation}
where $a$ and $b$ are the lengths of the major and minor axis, respectively. In the limit of small Capillary numbers (thus small deformations) a steady state is always achieved. In the limit of nearly spherical droplets an analytical solution was first reported in Ref. \cite{Shapira_Haber_1990} by solving the momentum and continuity equation using the Lorentz reflection method. Ref. \cite{Shapira_Haber_1990} gives the first order solution $D_{SH}$ (SH stands for the initials of the original authors in Ref. \cite{Shapira_Haber_1990}) for small confinements ratio $R/H$ between two parallel walls \cite{Shapira_Haber_1990,vananroye2007a}
\begin{equation}
D_{SH}= D_{Taylor} \left[1+C_{S H} \frac{1+2.5p}{1+p}\left(\frac{R}{ H}\right)^3\right],
\label{eq:deformation parameter}
\end{equation}
where $p$ is the viscosity ratio between the droplet and surrounding fluid, $H$ is the gap between the walls, $R$ is the droplet radius and parameter $C_{S H}$ which depends on the position of the droplet relative to the walls and this value grows when closer to the walls. Ref. \cite{Shapira_Haber_1990} performed analytical calculations for which they found that a droplet placed in the middle of the walls ($R/H=0.5$) has $C_{S H} = 5.699$ while a droplet at $R/H=0.125$ has $C_{S H} = 193.3$. Other values at different droplet positions are also reported. Some values have been tested against experiments as shown in Ref. \cite{vananroye2007a}.  The theoretical model given Eq. (\ref{eq:deformation parameter}) and Ref. \cite{Shapira_Haber_1990} has an additional term $D_{Taylor}$ obtained from Taylor small-deformation bulk theory \cite{taylor1932,taylor1934a} and is given by 
\begin{equation}
D_{Taylor} = \mathrm{Ca} \frac{16+19 p}{8(1+p)} \sin \theta \cos \theta,
\label{eq:deformationTaylor}
\end{equation}
where $\mathrm{Ca}$ is the Capillary number and $\theta$ is the angle between reference axis $x$ and major axis $a$ (see Fig. \ref{fig:capsuelanddropletinshear}) and usually assumed to be $\pi/4$ radians as in Ref. \cite{taylor1932,vananroye2007a}. Taylor bulk theory \cite{taylor1932,taylor1934a} is valid for small deformations and viscosity ratios $p$ close to 1. Furthermore, while this theoretical model predicts that the magnitude of the deformation will scale with the confinement ratio, the shape should remain ellipsoidal.

We highlight that this test can be performed in both 2D and 3D for droplets. For small $\mathrm{Ca}$ the deformation parameter is the same for a 2D and 3D droplet (same parameters), while for higher $\mathrm{Ca}$ results for 2D and 3D droplets start to deviate. An extensive study of 2D vs 3D droplets can be found in Ref. \cite{Afkhami_Yue_Renardy_2009}. The methodology presented here is valid for all the fluid-fluid models mentioned in this review.

\subsubsection{Capsule}
A capsule in shear will also reach a steady state and the main parameter defining the deformation is also the Capillary number now defined as $\mathrm{Ca} = \frac{u_w R \mu}{k_s H}$ where $k_s$ is the stretching coefficient and has the same physical units as the surface tension $\Gamma$. The Taylor deformation parameter is also used to characterise capsules in shear flow. We highlight that the correct constitutive model needs to be used to compare with other literature or analytical results. Results and validation for Hookean constitutive laws can be found in Ref. \cite{Gou_Huang_Ruan_Fu_2018, Rahmat_Meng_Emerson_Wu_Barigou_Alexiadis_2021,Sui_Chew_Roy_Chen_Low_2007}. For Skalak models we refer the reader to Ref. \cite{kruger_efficient_2011,Wouters_Aouane_Kruger_Harting_2019,Guglietta_Behr_Biferale_Falcucci_Sbragaglia_2020}. It is not trivial to obtain analytical results from constitutive models. In most cases, models are validated against other numerical implementations. An analytical result for the Taylor deformation parameter has been found for the neo-Hookean and Skalak laws, Eqs (\ref{eq:neohookean}) and (\ref{eq:skalak}) and is given by \cite{kruger_efficient_2011,Barthes-Biesel_1980,Barthes-Biesel_Rallison_1981},
\begin{equation}
D=\frac{5}{4} \frac{3 \alpha_2+4 \alpha_3}{\alpha_1\left(3 \alpha_2+5 \alpha_3\right)+2 \alpha_3\left(\alpha_2+\alpha_3\right)} \mathrm{Ca}+O\left(\mathrm{Ca}^2\right)
\label{eq:capsule_taylor_analyticaç}
\end{equation}
where $\alpha_1=0, \alpha_2=2 / 3$, and $\alpha_3=1 / 3$ are coefficients extracted from the constitutive model. Equation (\ref{eq:capsule_taylor_analyticaç}) is only valid for small deformation in Stokes flow ($\mathrm{Re}<1$). More details can be found in Ref. \cite{kruger_efficient_2011}

\subsection{Capsule sedimentation}
The process of capsule sedimentation involves placing a capsule in a static fluid and allowing it to accelerate downward under gravity until it reaches a terminal velocity, where the resultant drag force balances the gravitational load (see Fig. \ref{fig:diagram_and_terminalvelocity}). We present a simple validation for a confined 2D capsule using the harmonic potential given in Eq. (\ref{eq:harmonic_potential}). The capsule should be rigid. The elastic and bending coefficients can be adjusted to achieve a quasi-rigid state. It is clear that simulating sedimentation of a 2D capsule i.e. a cylinder, will encounter Stokes’ paradox, which states that there is no drag force on the cylinder in the Stokes regime. However, experiments and approximations have been derived for cylinder sedimentation in the presence of walls. Therefore, this benchmark consists of placing the 2D capsule between 2 vertical parallel walls. The top and bottom boundaries can be walls or periodic but the vertical walls need to be long enough to minimize boundary effects. 

According to Ref. \cite{ghosh_stockie_2015}, the analytical wall-corrected settling velocity of a 2D particle can be calculated from
\begin{equation}
V_{c}=\frac{\pi g D^{2}\left(\rho_{s}-\rho_{f}\right)}{4 \mu \Lambda},
\label{eqn:v_c}
\end{equation}
where $D$ is the diameter of the particle, $g$ is gravitational acceleration, $\rho$ is the density where $s$ and $f$ stand for solid and fluid, respectively, $\mu$ is the dynamic viscosity and $\Lambda$ is the wall correction factor. Subscript $c$ stands for corrected. Notice that Eq. (\ref{eqn:v_c}) is obtained by equating the gravitational force per unit length ${F}_g=\frac{\pi}{4} g D^2\left(\rho_s-\rho_f\right)$ with the drag force per unit length ${F}_d=V_c \mu \Lambda$. Moreover, Eq. (\ref{eqn:v_c}) is valid for low Reynolds numbers. There are several approximations for $\Lambda$ which can be found in Ref. \cite{ghosh_stockie_2015,ben_richou_drag_2005}. A commonly cited approximation can be found in Ref. \cite{faxen_1946,happel_low_1983} where Stokes equation for Newtonian fluids using asymptotic expansion with no-slip boundary conditions on the walls given by
\begin{equation}
\Lambda = \frac{-4 \pi}{0.9157+\ln (k)-1.724 k^{2}+1.730 k^{4}-2.406 k^{6}+4.591 k^{8}}
\label{eq:faxen}
\end{equation}
where $k=D/H$ is the dimensionless particle size. $D$ is the diameter of the particle and $H$ is the distance between the two vertical walls i.e., the width of the channel. Ref. \cite{faxen_1946,happel_low_1983} applied an integral transform which produces a convergent series and allows the boundary conditions to be applied simultaneously. It is generally known that Eq. (\ref{eq:faxen}) is valid for $k \in[0,0.5]$. Usually, $\Lambda$ is only a function of $k$ for low or very high Reynolds numbers $\mathrm{Re}$. For intermediate values, it is also a function of the Reynolds number. The results for a 2D sedimenting capsule are shown in Fig. \ref{fig:diagram_and_terminalvelocity} and are in agreement up to $k \approx 0.4$ in the range $k \in[0,0.5]$. Note that for a 3D spherical capsule, Stokes' law can be used for validation.

\begin{figure}
\center
\includegraphics[width=0.45\textwidth]{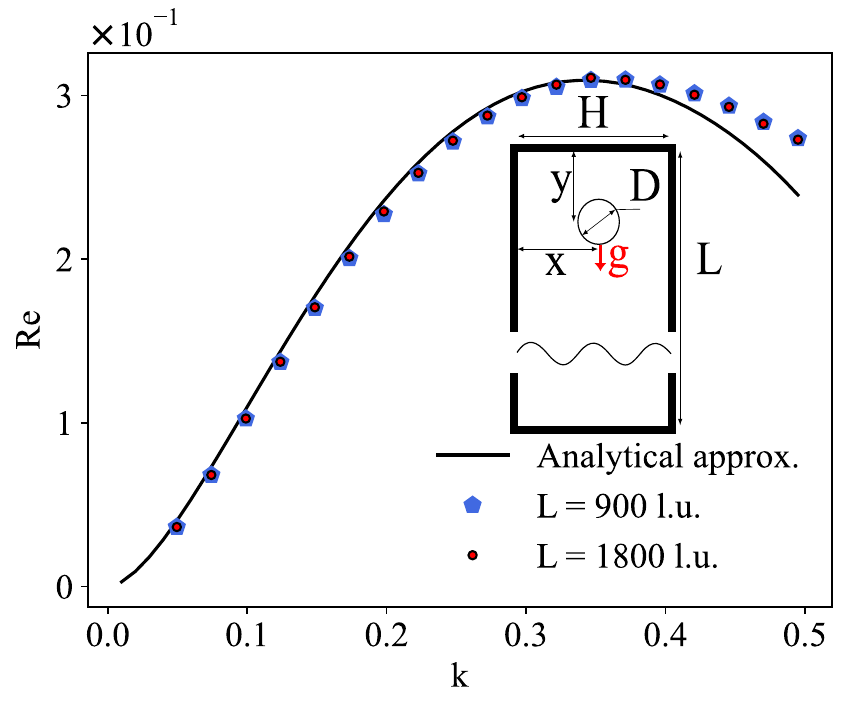}
\caption{(a) Schematic of an initially circular particle of diameter $D$ moving in a channel of length $L$ and width $H$ due to the force of gravity $\boldsymbol{g}$. The particle is placed at $x_0=0.5H$. (b) Terminal velocity $V_t$ against dimensionless size $k$. The particle has a diameter of 12.5 l.u and the fluid has a relaxation parameter $\tau = 1$. The analytical approximation corresponds to Eq. (\ref{eqn:v_c}).}
\label{fig:diagram_and_terminalvelocity}
\end{figure}

\subsection{Droplet Laplace test}
\label{subsec:laplacetestdroplet}
The Laplace test consists of placing a spherical droplet in the centre of a fluid domain and verifying the Laplace law. Laplace law dictates that when the system reaches equilibrium, the pressure difference across the droplet interface $\Delta P$ is related to the interfacial tension $\Gamma$  

\begin{equation}
    \Delta P = \frac{2\Gamma}{R},
\label{eq:laplace_law}
\end{equation}
where $R$ is the radius of the droplet. In 2D the Laplace pressure is $  \Delta P = \frac{\Gamma}{R}$. One should measure the pressure at the droplet centre and then as far away from the centre as possible to prevent undesired effects from affecting the measurements such as spurious velocities. The domain needs to be large enough to prevent the effect of periodic copies (periodic boundaries) or walls (solid boundaries). A linear relationship should be obtained and the slope is the surface tension. As an example, we plot results for a droplet using the pseudopotential method depicted in Fig. \ref{fig:laplace}. Additionally, it is possible to compare the surface tension value to theoretical and numerical values \cite{Montella_Chareyre_Salager_Gens_2020}. Moreover, other hydrostatic tests such as contact angle test \cite{Montella_Chareyre_Salager_Gens_2020} (for wetting boundary conditions) can be useful to further check the accuracy of the model. 

\begin{figure}
\center
\includegraphics[width=0.45\textwidth]{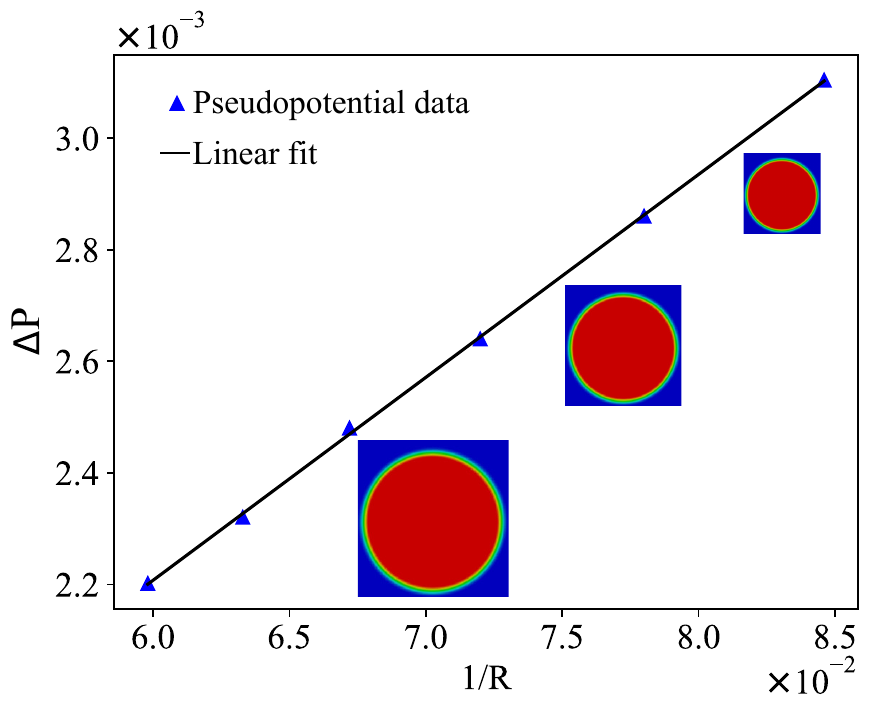}
\caption{Laplace test using the pseudopotential method at equilibrium for $G_{k, 1} = -7.9$, $G_{k, 2} = 4.9$ (see Eq. (\ref{eq:competing_force})) and $\tau_A=\tau_B=1$. From the linear fit we obtain the surface tension $\Gamma = 0.00367$. The images show the steady state of droplets as we vary the radius.}
\label{fig:laplace}
\end{figure}

\section{Summary and outlook}
\label{sec:conclusions}
We reviewed different methods to model fluid-filled objects in flow and 
addressed some of the advantages, disadvantages, and trade-offs of the different methodologies, as perceived based on our own (necessarily biased) experience and the literature survey. Below, we summarise the main items.

Fluid-fluid models are usually computationally more economical than fluid-structure ones because the interfaces emerge spontaneously from the underlying mesoscale physics and do not require any explicit tracking. The price for this flexibility is a lack of specific control of the mechanical properties of the interfaces, as often required in biological applications. They are also generally affected by spurious currents resulting from lack of sufficient isotropy of the underlying lattice; even though several efficient remedies have been developed over the years, spurious currents remain an issue to be critically monitored. 

Both pseudo-potential and free energy must be supplemented with a model for near-contact 
interactions in order to prevent droplet coalescence, especially under conditions 
typical of microfluidic experiments. 
Such near-contact interactions can be implemented very effectively within 
high-performance LB codes, yielding excellent performances on parallel GPU clusters. 
For instance,  optimized codes running on GPU accelerators can reach GLUPS
performance (one billion lattice updates per second) on single GPU machines with a few
thousands CUDA cores \cite{BONA22,BONA23}.
To convey a concrete idea means that one can simulate one millimeter 
cube material at micron resolution in space and nanosecond resolution in time, over 
one millisecond timespan in about $10^6$ seconds, namely about two weeks wall clock time. 
On massively parallel GPU clusters, these figures can be boosted
by two orders of magnitude, meaning that the above simulation can be 
performed in just a few hours of wall-clock time.  
These numbers open up exciting perspectives for the computational design of new
soft materials \cite{COPMAT,OUP18}.
Fluid-fluid models readily extend to three dimensions by simply enlarging the set of discrete
velocities, although at a correspondingly increased computational cost (especially
on account of memory access) \cite{Falcucci2021}.  

As mentioned above, fluid-structure models are computationally intensive but offer
a more accurate control of the mechanical properties of the interfaces, hence of the shape
of the resulting droplets and capsules.
They also provide a significant latitude in shape-space and do not suffer 
from significant spurious current effects. 
Since the structural dynamics is explicitly taken into account, the computational cost
raises with the number of droplets and the number of degrees of freedom per droplet/capsule,
setting the case for a judicious trade-off between physical fidelity and computational viability.   
Significant progress over the recent years has led to near-GLUPS performance on multi-million
grid sizes with embedded propellers \cite{LATT}.

Finally, three-dimensional extensions are demanding not only in terms of computational resources, but
also in terms of model and programming complexity, such as the switch to finite-element representations
of the internal degrees of freedom. 

By and large, it appears fair to say that the major progress of LB models over the last decade, both in its fluid-fluid and fluid-structure versions has come to a point of enabling
the computational study of new smart materials, both passive and active. 
Furthermore, the capability of tracking droplets and capsules in realistically complex geometries, such
as the human body, may disclose new exciting perspectives in computational microphysiology and medicine~\cite{coveney2023virtual}.

\section*{Acknowledgements}

We acknowledge financial support from the Portuguese Foundation for Science and Technology (FCT) under the contracts: PTDC/FIS-MAC/28146/2017 (LISBOA-01-0145-FEDER-028146), EXPL/FIS-MAC/0406/2021, UIDB/00618/2020, UIDP/00618/2020 and 2020.08525.BD. 
SS wishes to acknowledge financial support from the ERC-PoC grants DROPTRACK (contract number 101081171).

\bibliographystyle{rsc}
\bibliography{aipsamp}

\end{document}